# Hydrogen-Mediated Control of Phase Formation and Microstructure Evolution

**Lukas Schweiger**[a] *, **Florian Spieckermann**[a] **, Peter Cengeri[b], Alexander Schökel[c], Michael Zehetbauer[b], Erhard Schafler[b], Daniel Kiener[a], Jürgen Eckert[a,d]

[a] Department of Materials Science, Montanuniversität Leoben, Jahnstraße 12, 8700 Leoben, Austria

[b] Faculty of Physics, University of Vienna, 1090 Vienna, Austria

[c] Deutsches Elektronen-Synchrotron DESY, Notkestr. 85, 22607 Hamburg, Germany

[d] Erich Schmid Institute of Materials Science, Austrian Academy of Sciences, Jahnstraße 12, 8700 Leoben, Austria

* Corresponding authors: lukas.schweiger@unileoben.ac.at, florian.spieckermann@unileoben.ac.at

Abstract

Hydrogen is key in reducing greenhouse gas emissions in materials production. At the same time, it significantly affects mechanical properties, often causing unwanted embrittlement. However, rather than solely addressing these disadvantages, hydrogen's inevitable role in sustainable metallurgy should be leveraged to create new and potentially superior materials. Here, we show that using hydrogen in the form of metal hydrides introduces a barrier to mechanical alloying, stabilizing otherwise unattainable microstructures. Severe plastic deformation of a composite of the high entropy alloy (HEA) TiVZrNbHf and Cu leads to amorphization while substituting the HEA by its hydride preserves the two-phase structure. Monte Carlo simulations confirm that the significantly different hydrogen affinities, together with the restricted dislocation motion in the hydride, create a barrier to mechanical alloying. This hydride route enables new microstructural states, even in well-studied material systems. It opens an additional dimension in designing materials with diverging hydrogen affinities, offering tighter control over mechanical alloying.

High entropy alloys (HEAs) are intensely investigated for mechanical and functional properties.[1,2] In particular, hydride-forming HEAs are promising candidates for solid-state hydrogen storage due to their potential for tailoring thermodynamic properties for specific applications.[3]

Sahlberg *et al*. suggested that in the HEA TiVZrNbHf, due to the high lattice strain introduced by the differences in atomic radii, otherwise inaccessible interstitial sites become available to hydrogen, boosting the hydrogen storage capacity of the material.[4] Although subsequent investigations did not confirm such a lattice strain effect[5], TiVZrNbHf and similar variants remain among the best-researched HEA systems, particularly for hydrogen storage.[5–7] Therefore, this system was also chosen as a starting material in this investigation.

Besides developing new alloys, thermomechanical processing is also investigated to pave the way for further property improvements of HEAs:[8–10] This includes severe plastic deformation (SPD) methods, such as high-pressure torsion (HPT), which result in nanocrystalline HEAs with improved hydrogen sorption properties[11] or superplasticity[9].

In this context, it is essential to note that thermomechanical processing can not only enhance hydrogen sorption properties but vice versa, hydrogen and hydrides can significantly impact the plasticity and recovery/recrystallization of alloys.[12,13] For example, enhanced strength and ductility were reported for MnCrFeCoNi HEAs due to the impact of hydrogen on the stacking fault energy, enabling nano-twinning.[12] Other works reported accelerated grain growth induced by hydrogen in single-phase vanadium.[13] In contrast to these results, the authors have shown in a previous work that although HPT-deformed TiVZrNbHf is unstable under high-temperature conditions, hydrogen tends to stabilize its nanocrystalline structure.[14] Such a stabilization could also be associated with hydrogen-assisted spinodal decomposition, as recently observed by Ma et al.[15] This highlights that hydrogen can impact the microstructure and phase evolution of materials in manifold ways.

Consequently, hydrogen might be used as a temporary alloying element,[16] such as in Ti alloys during thermo-hydrogen processing[17] or for producing high-damping shape-memory alloys,[18] or hetero-structured materials by nano-twin gradients.[19] Further, with the projected widespread use of hydrogen for producing sustainable alloys, e.g., green steels[20,21], such processing routes will become more viable or even unavoidable.[22] Although hydrides are known for their brittle behavior[23], recent examples highlight that hydrides might enable overcoming strength-ductility trade-offs.[24] Beyond that, hydrides are particularly appealing due to their functional properties, such as superconductivity[25] and hydrogen storage.[26]

In this study, we push the concept of using hydrogen to tailor microstructures to the limits by applying it to composites of phases with widely diverging hydrogen affinities. Powders of the TiVZrNbHf HEA or the respective TiVZrNbHfH$_{\approx 10}$ HEA hydride are blended with Cu powder and subjected to HPT. Changing from metal to hydride leads to a drastically different microstructure evolution during plastic deformation. For the metal, a metallic glass is formed by mechanical alloying, while for the metal hydride, a two-phase nanocomposite is stabilized. The prospect of using hydrogen to inhibit mechanical alloying and create otherwise unattainable composite structures is highly appealing. It would be widely applicable to different alloy systems, particularly in view of the anticipated widespread use of hydrogen in future metal and materials production.

**Figure 1** shows backscatter electron (BSE) SEM micrographs of the TiVZrNbHf-Cu (HEA-Cu) and TiVZrNbHfH$_{\approx 10}$-Cu (HEA hydride-Cu) composites containing 53wt.% Cu, with all micrographs taken at a radius of 3 mm. The Cu phase appears black, and the HEA and HEA hydride bright. Additionally, the hydride samples exhibit some charging during SEM, likely due to the reduced electric conductivity of the hydride.

Both composites exhibit refinement of the HEA and HEA hydride phases, respectively. Nevertheless, as seen in **Figure 1 (b-i)**, differences in the microstructure evolution are visible already at low strains and become more apparent at higher strains.

At *n*=10 ($\gamma\approx$428/471), the HEA phase elongated drastically, while the hydride deformed to a smaller degree. The refinement appears to be more efficient in the HEA-Cu composite without hydrogen, and at *n*=50 ($\gamma\approx$2142), the HEA-Cu composite exhibits a uniform microstructure. No distinction between individual phases was possible, indicating that a nanocrystalline composite or a solid solution formed. Contrarily, the HEA hydride-Cu composite consists of two well-distinguishable but significantly refined phases with particle sizes mostly ≤ 1 μm. This is also confirmed by EDX maps in **Figure 1 (j,k)**, which clearly show the retained chemical partitioning in the hydride composite. Image segmentation results in **Figure 1 (l,m)** highlight the lower fraction of distinct HEA phase compared to the HEA hydride, the former being hardly resolvable at *n*=50 ($\gamma\approx$2142). A more detailed evaluation, including EDX, is given in the supplementary information in **Figure S2** and **Tables S1-2**.

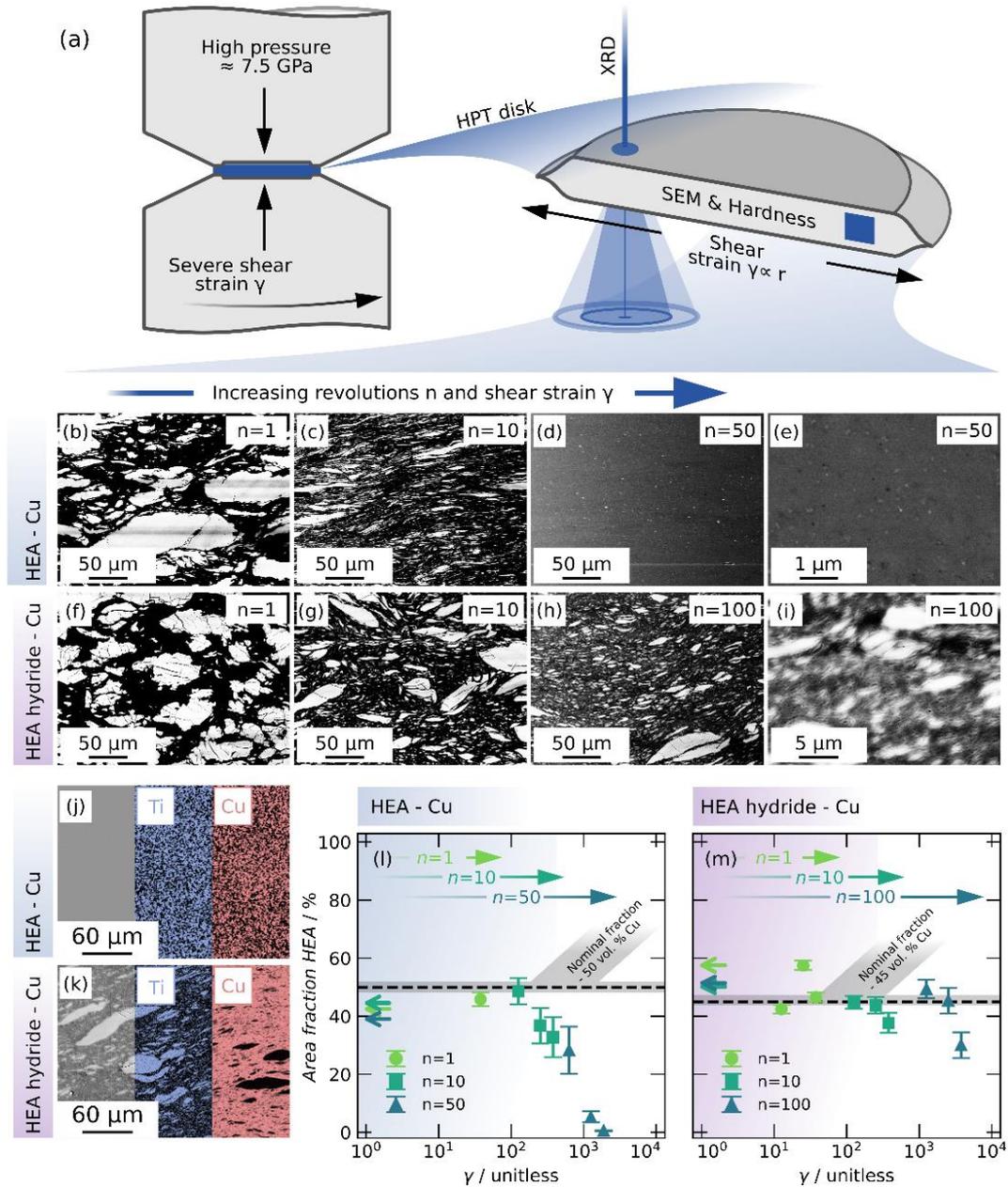

**Figure 1:** (a) Illustration of the HPT process and resulting disks with a radius-dependent shear strain $\gamma$. BSE SEM micrographs of HEA-Cu composites after HPT at RT and (b) $n$=1 ($\gamma$≈34), (c) $n$=10 ($\gamma$≈428) and (d,e) $n$=50 ($\gamma$≈2142) and of HEA hydride-Cu composites after HPT at RT and (f) $n$=1 ($\gamma$≈42), (g) $n$=10 ($\gamma$≈471) and (h, i) $n$=100 ($\gamma$≈4712). EDX maps at $n$=50 and 100 of the (j) HEA-Cu and (k) HEA hydride-Cu composites, respectively. All shown micrographs were taken at a radius of 3 mm. (l) HEA and (m) HEA hydride phase fractions were determined by image segmentation from the SEM micrographs. Arrows indicate the results at $r ≈ 0$ mm and therefore $\gamma ≈ 0$.

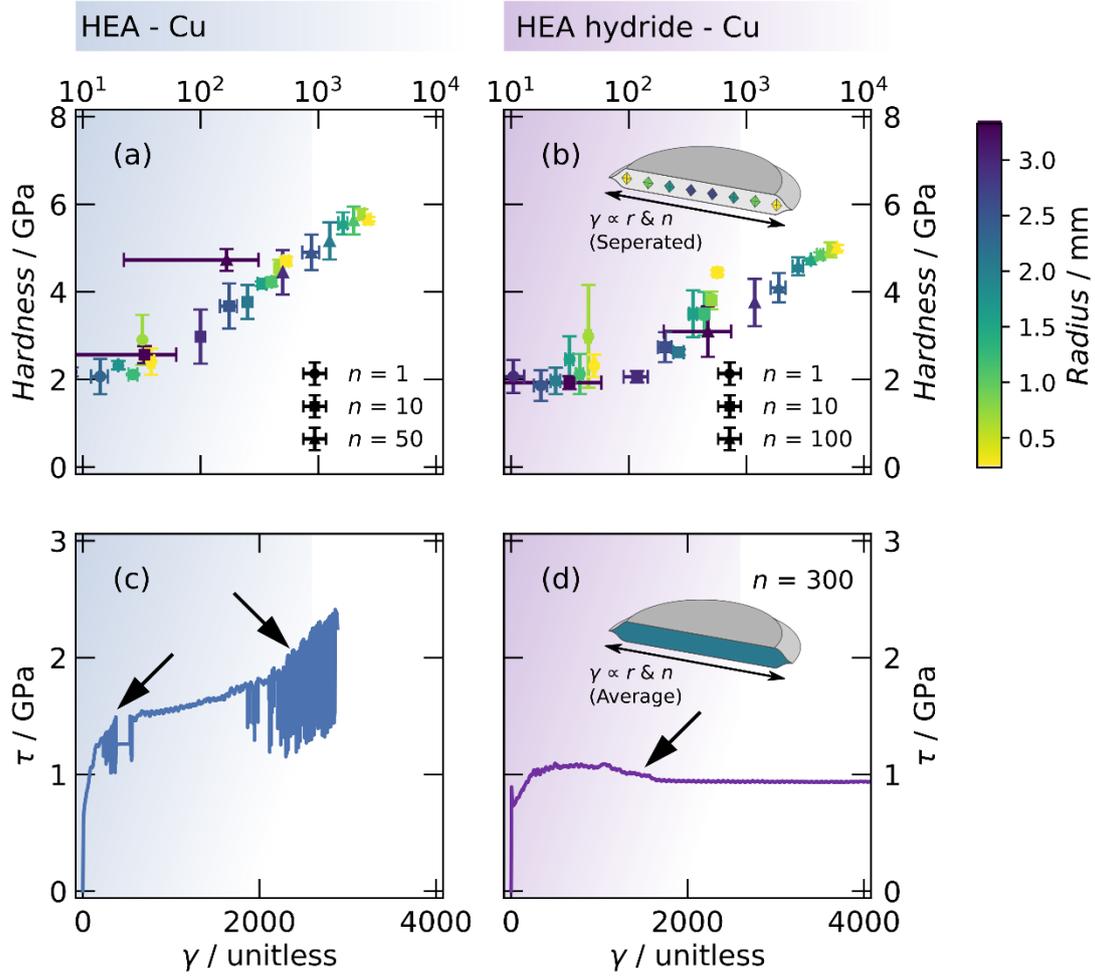

**Figure 2:** The hardness plotted as a function of the introduced shear strain $\gamma$ for (a) the HEA-Cu and (b) the HEA hydride-Cu composite. Flow curves constructed from torque measurements, which give the shear stress $\tau$ as a function of shear strain $\gamma$, are shown in (c) for the HEA-Cu and (d) for the HEA hydride-Cu composite.

The evolution of the mechanical properties during HPT can be followed (*ex-situ*) by measuring the radius-dependent hardness of the composite after HPT and by constructing flow curves from the torque recorded (*in situ*) during HPT deformation. Beyond basic mechanical properties, the former offers insight into the refinement of the composite, while the latter provides an impression of the deformation behavior of the material during HPT.

The hardness and flow curves of the HEA-Cu composite are plotted in **Figure 2 (a,c).** The hardness starts at the same value as for the hydride composite (**Figure 2 (b,d)**) and increases with HPT deformation, reaching nearly 6 GPa without any observable saturation plateau.

In line with this, the HEA-Cu composite exhibits a continuous increase in shear stress, indicating ongoing hardening without any signs of strain localization. One reason for this might lie in the better deformability of the metallic bcc HEA compared to the corresponding hydride. The differences in the deformation behavior of both phases are apparent in the SEM micrographs in **Figure 1**, with the HEA particles exhibiting significantly more elongation and necking. Localization events are likely suppressed by mechanical alloying and the resulting solid solution strengthening. The significant hardening of the HEA-Cu composites limits further deformation as plastic instabilities or slip-stick events, indicated by arrows in **Figure 2 (c,d),** occur at higher numbers of revolutions.[29] In the latter case, the contact friction between anvil and sample becomes too weak to facilitate further HPT processing and refinement.

The hardness of the HEA hydride-Cu composite is plotted against the introduced strain in **Figure 2 (b),** revealing an increase in hardness from ≈ 2 GPa to ≈ 5 GPa, and a plateau at higher strains.

The complementary **Figure 2 (d)** depicts the flow curve of the HEA hydride-Cu composite. The curve exhibits an initial rise in shear stress $\tau$, followed by a slight reduction, presumably associated with the onset of strain localization in the softer composite phase, i.e., in Cu. Such strain localization significantly imparts further refinement of the HEA hydride phase, for example, observed in the Mg-Fe system.[30] Higher resolution micrographs in **Figure 1** show limited refinement; no homogenous nanocomposite, as for the W-Cu[31,32] or FeTi-Cu[27,33] systems, was obtained. Increasing the number of revolutions further to $n$=300 results not in further refinement but severe cracking.

Synchrotron XRD measurements in **Figure 3** confirm the significant structural differences after HPT. Both HEA-Cu and HEA hydride-Cu composites initially consist of a distinct two-phase microstructure. At higher strains, the HEA-Cu composites transform into a single-phase material and exhibit (partial) amorphization, as highlighted by the broad diffraction maxima in the XRD patterns. A minor shoulder associated with the HEA is still visible, highlighting the incomplete amorphization. Further XRD patterns plotted in **Figure S3** and recorded as a function of radius show that amorphization only occurs at the largest radii due to the largest strains being present here. In fact, the HEA-Cu composite includes elements like Ti, Zr, and Cu, which are known to promote amorphization so that partial amorphization can be expected.[34] Studies on Ti-(Zr)-Cu also found partial amorphization during HPT.[35,36]

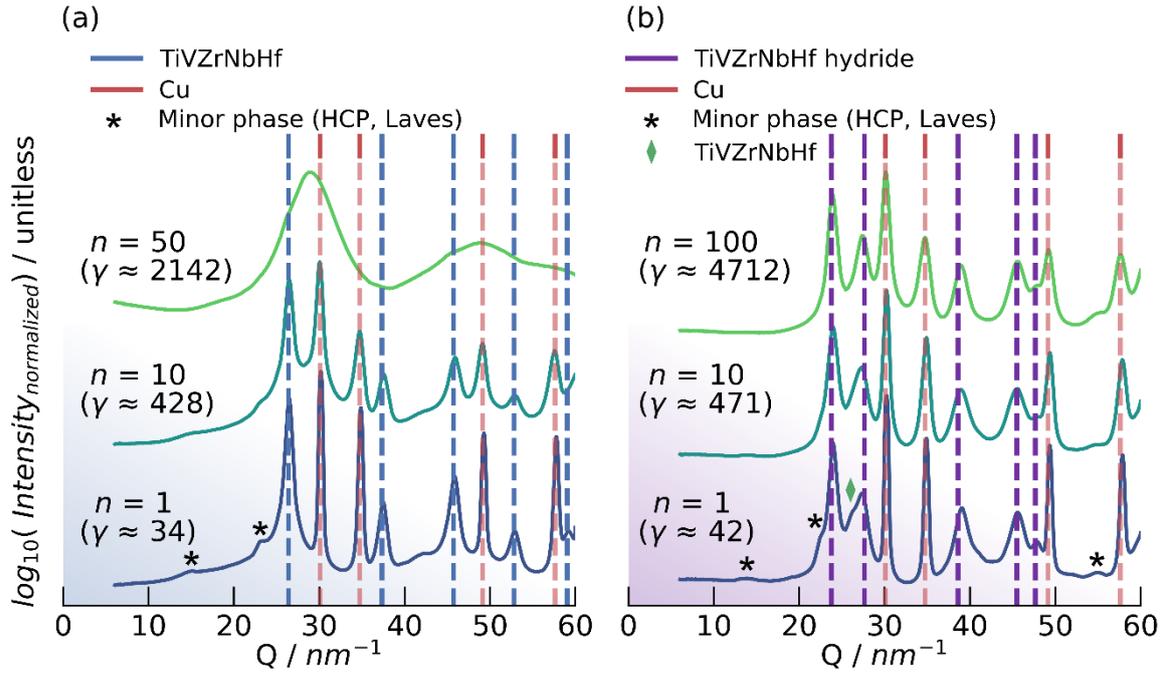

**Figure 3:** Synchrotron XRD patterns of (a) the HEA-Cu and (b) the HEA hydride-Cu composites at $n$=1, 10, 50, and 100, respectively. The beam was centered at $r \approx 3mm$. The corresponding shear strains $\gamma$ at these radii are given.

In agreement with the microstructure determined by SEM and irrespective of strain, the HEA hydride-Cu composite retains the two-phase structure, with X-ray peak broadening due to the grain refinement and dislocation accumulation in both phases during HPT deformation. This indicates that hydrogen significantly restricts mechanical alloying.

DSC measurements revealed distinct exothermic crystallization peaks in the case of the (partially) amorphous HEA-Cu composites. The most severely deformed HEA hydride-Cu composite decomposes exothermically immediately after dehydrogenation. These results, detailed in **Figures S4-6** of the supplementary information, confirm the importance of hydrogen in stabilizing the otherwise unstable two-phase microstructure.

The results show that replacing the metal with its corresponding hydride effectively inhibits mechanical alloying. Mechanical alloying occurs via mechanical mixing and diffusion-assisted mechanisms.[37] Mechanical mixing models, such as kinetic roughening[38] and dislocation shuffling[39], involve continuous shearing of the respective phases, irrespective of mixing enthalpies. Pairings with unlikely mechanical properties, such as HEA hydride-Cu, tend to follow an erosion and abrasion model, leading to gradual refinement and subsequent mechanical mixing.[33,40,41]

Diffusion can aid or hinder alloying, depending on the mixing enthalpy.[42] However, even for a positive enthalpy of mixing, high defect and phase boundary densities can drive diffusion-assisted dissolution, with small particles dissolving due to the Gibbs-Thomson effect.[39]

These mechanisms are not necessarily exclusive but can work sequentially, with initial mechanical refinement, e.g., by dislocation shuffling, enabling subsequent atomic-scale dissolution by diffusion.[39] Consequently, the observed suppression of mechanical alloying could originate from both limited diffusion and/or mechanical mixing.

Regarding diffusion, the impeded mechanical alloying and related composite stabilization are connected to the different affinities of the respective composite phases for hydrogen. The HEA, consisting of elements with high hydrogen affinity, has negative enthalpies of hydrogen dissolution and hydride formation. For metallic Cu, in contrast, the enthalpy of dissolution is positive.[6,43,44] The corresponding values are given in **Table 1**. Consequently, we hypothesize that the diverging structural evolution is rooted in the different hydrogen affinities and associated thermodynamics, particularly in the later stages of HPT. Therefore, the thermodynamics of the HEA(H)-Cu system is outlined below. For details, the reader is referred to the supplementary information in **Tables S3-4 and Equations S1-4.**

**Table 1:** Enthalpies of hydrogen dissolution $\Delta H_{diss}$ and enthalpies of hydride formation $\Delta H_{hydride}$ of Cu and the HEA elements.[6,43,44]

|  | **Cu**[43,44] | **Ti**[43,44] | **V**[43,44] | **Zr**[43,44] | **Nb**[43,44] | **Hf**[43,44] | **TiVZrNbHf**[6] |
|---|---|---|---|---|---|---|---|
| $\Delta H_{diss}$ / kJ mol$^{-1}$ H | + 43-55 | − 52 (α-Ti) | − 26-33 | −52-64 (α-Zr) | −33-38 | −38 | n.a. |
| $\Delta H_{hydride}$ / kJ mol$^{-1}$ H | n.a. | −68 (TiH$_2$) | −36-42 (VH$_{0.5}$) | −82-106 (ZrH$_2$) | −38 (NbH$_{0.5}$) | −66 (HfH$_2$) | −30 (MH$_{1.9-2.5}$) |

Without hydrogen, the constituent HEA atoms, i.e., Ti, V, Zr, Nb, Hf, will dissolve in the Cu phase during mechanical alloying. The enthalpies and entropies of mixing, $\Delta H_{mix}$ and $\Delta S_{mix}$, for a Cu-TiVZrNbHf phase were derived according to the procedure proposed by Yang and Zhang[45], using thermodynamic data from Takeuchi and Inoue[46]. The resultant free enthalpy $\Delta G_{mix}$ is found as −10.2 kJ mol$^{-1}$, and consequently, a single-phase solid-solution, either crystalline or amorphous, is expected, as confirmed experimentally.

However, in the case of the hydride, hydrogen must also be accommodated. Therefore, metal and hydrogen dissolution are coupled, and both hydrogen release from the HEA hydride and hydrogen dissolution into the Cu phase must occur. Assuming a hydrogen-to-metal ratio of two, two hydrogen atoms must be dissolved in Cu for every HEA atom.[6] The free enthalpy $\Delta G_{mix}$ for this process was derived by considering $\Delta H_{diss}$ and $\Delta H_{hydride}$ for Cu and HEA in **Table 1**, as well as the entropy gain by spreading hydrogen over both phases, yielding a $\Delta G_{mix}$-value of +17.5 or +21.2 kJ mol$^{-1}$ assuming a bcc or fcc solid solution, respectively. This coupling of metal and hydrogen diffusion imparts a significant positive enthalpy on the whole process, and the free enthalpy of mixing overall becomes $\gg 0$. Consequently, mechanical alloying and the formation of a single-phase solid solution and subsequent amorphization become thermodynamically blocked: *Diffusion will not support but counteract any mechanical mixing*. The underlying principle is illustrated in **Figure 4 (a)**.

The second critical factor is mechanical mixing itself, which is tightly linked to the required flow stresses, plastic deformation behavior, and the associated fragmentation of the constituent phases. In the HEA-Cu composite, the anticipated formation of an amorphous phase boundary between Cu and HEA promotes more uniform stress transmission across the interface, facilitating homogenous dislocation nucleation and subsequent co-deformation and mechanical mixing.[47,48]

In contrast, metal hydrides are generally considered brittle compared to their metallic counterparts. Limited plasticity and dislocation motion are possible in compression but are severely limited.[23] At the same time, cross-slip is also highly restricted.[23] Dislocation shuffling[39], i.e., the propagation of dislocations through phase boundaries, could thereby be prevented due to the change in crystal structure from bcc to distorted fcc or the lowered dislocation mobility. Therefore, such material pairing will likely inhibit co-deformation and mechanical mixing but promote deformation localization. Based on the flow curves in **Figure 2**, the latter was observed in the HEA hydride-Cu but not the HEA-Cu composite. As illustrated in **Figure 4 (b)**, this localization tendency is promoted by the absence of mechanical alloying and subsequent solid solution strengthening.

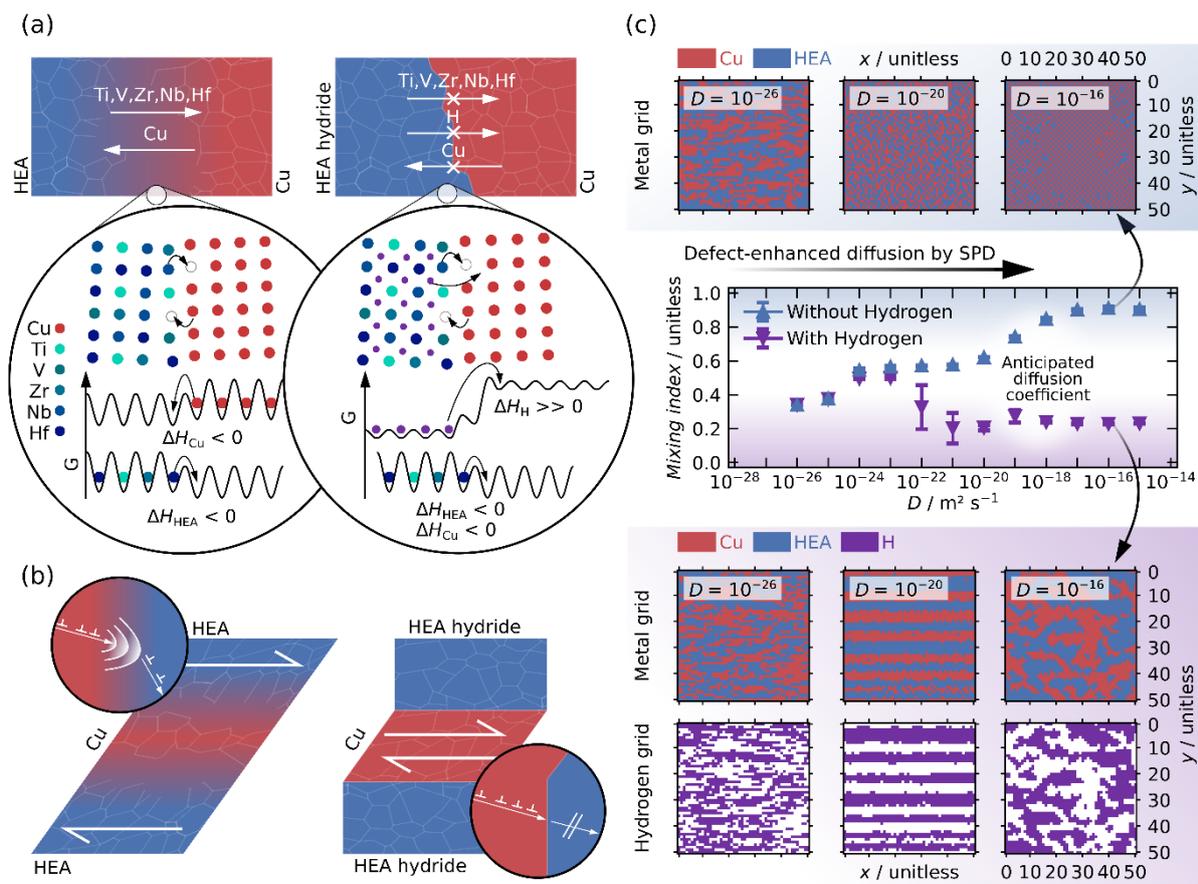

**Figure 4:** Illustration of the mechanisms impeding (a) mechanical alloying and (b) co-deformation during deformation of the HEA-Cu and HEA hydride-Cu composites. (c) Mixing indices derived from the Monte Carlo simulations assuming different diffusion coefficients including exemplary metal and hydrogen grids.

To support these considerations, Monte Carlo (MC) simulations were performed to mimic and conceptualize the microstructure evolution, using the enthalpies of mixing $\Delta H_{mix}$ estimated above as input parameters. Metal-metal composites and metal hydride-metal composites were modeled, including metal and hydrogen diffusion, and shear steps representing HPT-deformation were introduced. The latter was realized by translating one part of the grid by one site along a randomly chosen vertical or horizontal line. The respective probabilities are based on the experimental strain rates and assumed diffusion coefficients.

The results are summarized in **Figure 4 (c)**, and a more complete overview of the assumptions and results is provided in the supplementary information. The diffusion coefficient is crucial in setting the diffusion and shear probabilities during each MC step. Therefore, the diffusion coefficient was varied from $10^{-15}$ to $10^{-26}$ m$^2$ s$^{-1}$, as this input parameter needed to be estimated, and studies suggest significantly increased diffusion coefficients during SPD due to high defect densities.[49,50]

Comparing the results in **Figure 4 (c)**, one can see that despite the unchanged metal-metal interactions, the significant differences in the metal-hydrogen interactions result in an entirely different microstructure evolution. The results of the metal and hydride composites only converge at the lowest diffusion coefficients. Under such circumstances, mechanical mixing by shear, which neglects any associated energy changes, dominates, and diffusion hardly occurs. The MC simulations, however, do not include the restricted dislocation shuffling in the hydride case. More detailed results of the MC simulation are provided in the supplementary information, including **Figures S7-S14.**

As mentioned above, in the case of the HEA-Cu composite, mechanical alloying and the strengthening of the associated solid solution can potentially reduce differences in flow stress between the respective composite phases. Contrarily, the absence of appreciable mechanical alloying in the hydride composite will render localization or shear band formation more likely. The Monte Carlo simulations confirm these experimental observations. When the probability of the shear steps is scaled by a hypothetical flow stress derived from the chemical composition along the shear line, interdiffusion in the HEA-Cu composite tends to smoothen the localization, as seen in **Figure S8**. In the case of the HEA hydride-Cu composite, such localization phenomena are more pronounced, see **Figure S10,** even more so in the MC simulations introducing only horizontal shear. Therefore, the MC model confirms the experimental results of the flow curves in **Figure 2**, highlighting that hydrogen and its subsequent effects can significantly influence material behavior during severe plastic deformation, impacting microstructure evolution and coupled phase transformations.

Overall, the experimental and MC results align well, offering deeper insights into the connections between thermodynamics, deformation behavior, and resulting microstructure evolution of the system during SPD.

In this study, we used hydrogen to achieve microstructures otherwise not attainable by SPD. HPT was used to deform a composite composed of the TiVZrNbHf HEA and Cu, and due to the unavoidable mechanical alloying and interdiffusion, a (partially) amorphous material formed. However, exchanging the HEA with the corresponding TiVZrNBHfH$_{\approx 10}$ hydride enables an entirely different microstructure evolution. The proposed reason lies in the significantly different affinities of the constituent phases to hydrogen, with the HEA possessing a high and Cu having a low hydrogen affinity. Because significant mechanical alloying of metal atoms also requires the redistribution of hydrogen, these processes become energetically coupled, and consequently, a sizeable energy barrier against mechanical alloying is introduced. In this work, this mechanism was confirmed by a simplified model based on Monte Carlo simulations, mimicking the chemically complex system. In addition, the absence of mechanical alloying and restricted dislocation mobility in the hydride limits the direct mechanical mixing of the constituent phases.

It is concluded that a *hydrogen-induced barrier for mechanical alloying* during HPT, as an exemplary deformation technique, can be introduced. The approach investigated in the present work paves the way for designing and producing complex multiphase composites, which are otherwise difficult to obtain but may possess unique functional properties, by utilizing hydrogen as a, potentially temporary, alloying element.

**Methods**

The HEA TiVZrNbHf was synthesized by arc melting (Arc Melter AM 0.5, Edmund Bühler GmbH) stoichiometric amounts of Ti (99.995%, HMW Hauner GmbH), V (99.9%), Zr (99.2%), Nb (99.9%), and Hf (99.9%). Subsequently, the ingot was exposed to 40 bar of hydrogen at 350 °C, causing self-pulverization due to the substantial volume expansion ($\approx 26\%$)[6] upon hydride formation.

The obtained hydride powder was heated to 500 °C in a vacuum furnace to desorb the hydrogen, yielding the corresponding dehydrogenated HEA powder. X-ray diffraction (XRD) patterns before and after annealing, as shown in **Figure S1**, confirm complete desorption. The resultant HEA and respective HEA hydride were blended with Cu powder to obtain composites with 53 wt.% Cu. The powders were mixed using a mortar and pestle in a glove box with $O_2$ and $H_2O$ levels below 0.5 ppm. The composites were subsequently compacted under inert gas conditions as described elsewhere.[27] Following, severe plastic deformation was introduced by HPT at room temperature (RT), a pressure of 7.5 GPa and under ambient atmosphere, with anvils having a cavity diameter and depth of 8 mm and 0.15 mm, respectively. The speed of deformation was set to 1.27 rpm. The anvils were cooled with pressurized air during deformation to avoid excessive heating, and the temperature is expected to remain significantly below the desorption temperature of the HEA hydride. The torque $T$ was measured during HPT to determine the torsional shear stress $\tau$ present during deformation, which was calculated by

$$\tau = \frac{3T}{2\pi r^3} \quad . \tag{1}$$

with $r$ being the radius of the HPT disk. The torsional shear strain $\gamma$ was calculated as

$$\gamma = \frac{2\pi n r}{t} \quad , \tag{2}$$

with $n$ being the number of revolutions, $r$ the radius at which the strain was calculated, and $t$ the thickness of the HPT disk.[28] The HPT disks had a diameter and thickness of about 8 mm and 0.5 mm, respectively.

The resultant microstructure was investigated by scanning electron microscopy (SEM; LEO type 1525, Carl Zeiss GmbH) and energy dispersive X-ray spectroscopy (EDX; XFlash 6–60, Bruker corporation) on mirror-polished cross-sections of the HPT disks. The same cross-section was subsequently used to determine the radius-dependent microhardness (DuraScan, ZwickRoell GmbH). X-ray diffraction (XRD) measurements were done at the P02.1 Powder Diffraction and Total Scattering Beam Line of PETRA III (DESY Hamburg) in transmission geometry, with a photon energy of 60 keV and a Dectris EIGER2X CdTe 1M-W detector. The calibration was done using a $CeO_2$ calibrant. The HEA and HEA hydride starting powders were characterized using a benchtop X-ray powder diffractometer using a Co-$K_\alpha$ source (D2 Phaser, Bruker).

Differential scanning calorimetry (DSC, heat flux differential scanning calorimeter Netzsch DSC 404 F1. Pegasus) was conducted on quartered HPT disks with the less deformed central region, i.e., $r < 2$ mm, removed.

Monte Carlo (MC) simulations were implemented using Python. The input parameters determining metal and hydrogen diffusion and the shear processes mimicking HPT deformation were based on reported thermodynamic data and the experimental strain rates. Details of the thermodynamic calculations and the simulations are provided in the supplementary information.

**Acknowledgments**

This research activity is part of the Strategic Core Research Area SCoRe A+ Hydrogen and Carbon and has received funding from Montanuniversität Leoben. We acknowledge DESY (Hamburg, Germany), a member of the Helmholtz Association HGF, for the provision of experimental facilities. Parts of this research were carried out at PETRA III using beamline P02.1. Beamtime was allocated for proposal I-20230216 EC. We acknowledge the support from Felix Römer and Eray Yüce as well as the good collaboration with Gökhan Gizer and Claudio Pistidda during this beamtime. We gratefully acknowledge the fruitful discussion with Claus Trost regarding the Monte Carlo simulations.

**Ethics declarations:**

**Competing interests:** The authors declare no competing interests

**Contributions**

**LS** prepared the samples, did microstructure characterization, analyzed the data, and wrote the original draft with input from **FS, DK. LS, FS** designed the experimental concept. **FS, DK, JE** supervised the project. **PC** performed the hydrogenation of the supplied HEA. **JE, ES, MZ** provided resources and methodology. **AS** supported the Synchrotron XRD measurements and provided methodology. **All authors** revised and edited the manuscript.

**Data availability.** The data that support this work are presented in the article and its Supplementary Information. Further data are available from the corresponding authors upon reasonable request.

**Code availability.** Analysis scripts for all data, as well as the Monte Carlo Code included in this work, are available from the corresponding authors upon reasonable request.

**References**

[1] E.P. George, D. Raabe, and R.O. Ritchie, "High-entropy alloys," Nat. Rev. Mater. **4**(8), 515–534 (2019).

[2] X. Wang, W. Guo, and Y. Fu, "High-entropy alloys: Emerging materials for advanced functional applications," J. Mater. Chem. A **9**(2), 663–701 (2021).

[3] F. Marques, M. Balcerzak, F. Winkelmann, G. Zepon, and M. Felderhoff, "Review and outlook on high-entropy alloys for hydrogen storage," Energy Environ. Sci. **14**(10), 5191–5227 (2021).

[4] M. Sahlberg, D. Karlsson, C. Zlotea, and U. Jansson, "Superior hydrogen storage in high entropy alloys," Sci. Rep. **6**, 1–6 (2016).

[5] M.M. Nygård, G. Ek, D. Karlsson, M. Sahlberg, M.H. Sørby, and B.C. Hauback, "Hydrogen storage in high-entropy alloys with varying degree of local lattice strain," Int. J. Hydrogen Energy **44**(55), 29140–29149 (2019).

[6] D. Karlsson, G. Ek, J. Cedervall, C. Zlotea, K.T. Møller, T.C. Hansen, J. Bednarčík, M. Paskevicius, M.H. Sørby, T.R. Jensen, U. Jansson, and M. Sahlberg, "Structure and Hydrogenation Properties of a HfNbTiVZr High-Entropy Alloy," Inorg. Chem. **57**(4), 2103–2110 (2018).

[7] M.M. Nygård, G. Ek, D. Karlsson, M.H. Sørby, M. Sahlberg, and B.C. Hauback, "Counting electrons - A new approach to tailor the hydrogen sorption properties of high-entropy alloys," Acta Mater. **175**, 121–129 (2019).

[8] B. Schuh, F. Mendez-Martin, B. Völker, E.P. George, H. Clemens, R. Pippan, and A. Hohenwarter, "Mechanical properties, microstructure and thermal stability of a nanocrystalline CoCrFeMnNi high-entropy alloy after severe plastic deformation," Acta Mater. **96**, 258–268 (2015).

[9] N.T.-C. Nguyen, P. Asghari-Rad, P. Sathiyamoorthi, A. Zargaran, C.S. Lee, and H.S. Kim, "Ultrahigh high-strain-rate superplasticity in a nanostructured high-entropy alloy," Nat. Commun. **11**(1), 2736 (2020).

[10] P. Shi, W. Ren, T. Zheng, Z. Ren, X. Hou, J. Peng, P. Hu, Y. Gao, Y. Zhong, and P.K. Liaw, "Enhanced strength–ductility synergy in ultrafine-grained eutectic high-entropy alloys by inheriting microstructural lamellae," Nat. Commun. **10**(1), 489 (2019).

[11] J. Hidalgo-Jimenez, J.M. Cubero-Sesin, K. Edalati, S. Khajavi, and J. Huot, "Effect of high-pressure torsion on first hydrogenation of Laves phase Ti0.5Zr0.5(Mn1-Fe )Cr1 (x = 0, 0.2 and 0.4) high entropy alloys," J. Alloys Compd. **969**(September), 172243 (2023).

[12] H. Luo, Z. Li, and D. Raabe, "Hydrogen enhances strength and ductility of an equiatomic high-entropy alloy," Sci. Rep. **7**(1), 1–7 (2017).

[13] M.L. Martin, A. Pundt, and R. Kirchheim, "Hydrogen-induced accelerated grain growth in vanadium," Acta Mater. **155**, 262–267 (2018).

[14] L. Schweiger, F. Römer, G. Gizer, M. Burtscher, D. Kiener, C. Pistidda, A. Schökel, F. Spieckermann, and J. Eckert, "Mechanical processing and thermal stability of the equiatomic high entropy alloy TiVZrNbHf under vacuum and hydrogen pressure," Appl. Phys. Lett. **124**(24), (2024).

[15] C. Wu, L.S. Aota, J. Rao, X. Zhang, L. Perrière, M.J. Duarte, D. Raabe, and Y. Ma, "Hydrogen-assisted spinodal decomposition in a TiNbZrHfTa complex concentrated alloy," Acta Mater. **285**(November 2024), 120707 (2025).

[16] N. Eliaz, D. Eliezer, and D.L. Olson, "Hydrogen-assisted processing of materials," Mater. Sci. Eng. A **289**(1), 41–53 (2000).

[17] O. Senkov, "Thermohydrogen processing of titanium alloys," Int. J. Hydrogen Energy **24**(6), 565–576 (1999).

[18] A. Biscarini, B. Coluzzi, G. Mazzolai, A. Tuissi, and F.M. Mazzolai, "Extraordinary high damping of hydrogen-doped NiTi and NiTiCu shape memory alloys," J. Alloys Compd. **355**(1–2), 52–57 (2003).

[19] H. Luo, W. Lu, X. Fang, D. Ponge, Z. Li, and D. Raabe, "Beating hydrogen with its own weapon: Nano-twin gradients enhance embrittlement resistance of a high-entropy alloy," Mater. Today **21**(10), 1003–1009 (2018).

[20] Ö. Özgün, X. Lu, Y. Ma, and D. Raabe, "How much hydrogen is in green steel?," Npj Mater. Degrad. **7**(1), 1–5 (2023).

[21] M. Jovičević-Klug, I.R. Souza Filho, H. Springer, C. Adam, and D. Raabe, "Green steel from red mud through climate-neutral hydrogen plasma reduction," Nature **625**(7996), 703–709 (2024).

[22] D. Raabe, "The Materials Science behind Sustainable Metals and Alloys," Chem. Rev. **123**(5), 2436–2608 (2023).

[23] H.K. Birnbaum, "Mechanical properties of metal hydrides," J. Less-Common Met. **104**(1), 31–41 (1984).

[24] L. Shao, Q. Luo, M. Zhang, L. Xue, J. Cui, Q. Yang, H. Ke, Y. Zhang, B. Shen, and W. Wang, "Dual-phase nano-glass-hydrides overcome the strength-ductility trade-off and magnetocaloric bottlenecks of rare earth based amorphous alloys," Nat. Commun. **15**(1), 1–8 (2024).

[25] G.S. Boebinger, A. V. Chubukov, I.R. Fisher, F.M. Grosche, P.J. Hirschfeld, S.R. Julian, B. Keimer, S.A. Kivelson, A.P. Mackenzie, Y. Maeno, J. Orenstein, B.J. Ramshaw, S. Sachdev, J. Schmalian, and M. Vojta, "Hydride superconductivity is here to stay," Nat. Rev. Phys. **7**(1), 2–3 (2024).

[26] R. Mohtadi, and S. Orimo, "The renaissance of hydrides as energy materials," Nat. Rev. Mater. **2**(3), 16091 (2016).

[27] L. Schweiger, D. Kiener, M. Burtscher, E. Schafler, G. Mori, F. Spieckermann, and J. Eckert, "From unlikely pairings to functional nanocomposites: FeTi–Cu as a model system," Mater. Today Adv. **20**(October), 100433 (2023).

[28] R. Pippan, in *Bulk Nanostructured Mater.*, edited by M.J. Zehetbauer and Y.T. Zhu (Wiley-VCH Verlag GmbH & Co. KGaA., Weinheim, 2009), pp. 217–234.

[29] A. Hohenwarter, A. Bachmaier, B. Gludovatz, S. Scheriau, and R. Pippan, "Technical parameters affecting grain refinement by high pressure torsion," Int. J. Mater. Res. **100**(12), 1653–1661 (2009).

[30] M. Roostaei, P.J. Uggowitzer, R. Pippan, and O. Renk, "Severe plastic deformation close to the melting point enables Mg-Fe nanocomposites with exceptional strength," Scr. Mater. **230**(March), 115428 (2023).

[31] M. Zhao, I. Issa, M.J. Pfeifenberger, M. Wurmshuber, and D. Kiener, "Tailoring ultra-strong nanocrystalline tungsten nanofoams by reverse phase dissolution," Acta Mater. **182**, 215–225 (2020).

[32] M. Wurmshuber, M. Burtscher, S. Doppermann, R. Bodlos, D. Scheiber, L. Romaner, and D. Kiener, "Mechanical performance of doped W–Cu nanocomposites," Mater. Sci. Eng. A **857**(September), 144102 (2022).

[33] L. Schweiger, F. Spieckermann, N. Buchebner, J.F. Keckes, D. Kiener, and J. Eckert, "Exploring Refinement Characteristics in FeTi–Cu x Composites: A Study of Localization and Abrasion Constraints," Adv. Eng. Mater. **26**(19), (2024).

[34] P. Gong, L. Deng, J. Jin, S. Wang, X. Wang, and K. Yao, "Review on the Research and Development of Ti-Based Bulk Metallic Glasses," Metals (Basel). **6**(11), 264 (2016).

[35] Á. Révész, S. Hóbor, J.L. Lábár, A.P. Zhilyaev, and Z. Kovács, "Partial amorphization of a Cu–Zr–Ti alloy by high pressure torsion," J. Appl. Phys. **100**(10), (2006).

[36] Y.F. Sun, H. Fujii, T. Nakamura, N. Tsuji, D. Todaka, and M. Umemoto, "Critical strain for mechanical alloying of Cu–Ag, Cu–Ni and Cu–Zr by high-pressure torsion," Scr. Mater. **65**(6), 489–492 (2011).

[37] K.S. Kormout, R. Pippan, and A. Bachmaier, "Deformation-Induced Supersaturation in Immiscible Material Systems during High-Pressure Torsion," Adv. Eng. Mater. **19**(4), 1600675 (2017).

[38] P. Bellon, and R.S. Averback, "Nonequilibrium roughening of interfaces in crystals under shear: Application to ball milling," Phys. Rev. Lett. **74**(10), 1819–1822 (1995).

[39] D. Raabe, S. Ohsaki, and K. Hono, "Mechanical alloying and amorphization in Cu-Nb-Ag in situ composite wires studied by transmission electron microscopy and atom probe tomography," Acta Mater. **57**(17), 5254–5263 (2009).

[40] A. Bachmaier, G.B. Rathmayr, M. Bartosik, D. Apel, Z. Zhang, and R. Pippan, "New insights on the formation of supersaturated solid solutions in the Cu-Cr system deformed by high-pressure torsion," Acta Mater. **69**, 301–313 (2014).

[41] A. Bachmaier, J. Schmauch, H. Aboulfadl, A. Verch, and C. Motz, "On the process of co-deformation and phase dissolution in a hard-soft immiscible Cu Co alloy system during high-pressure torsion deformation," Acta Mater. **115**, 333–346 (2016).

[42] M.A. Shtremel, "Participation of diffusion in the processes of mechanical alloying," Met. Sci. Heat Treat. **44**(7–8), 324–327 (2002).

[43] R. Griessen, and T. Riesterer, in *Hydrog. Intermet. Compd. I*, edited by L. Schlapbach (Springer-Verlag Berlin Heidelberg, Berlin, Heidelberg, 1988), pp. 219–284.

[44] Y. Fukai, *The Metal-Hydrogen System*, 2nd ed. (Springer Berlin Heidelberg, Berlin, Heidelberg, 2005).

[45] X. Yang, and Y. Zhang, "Prediction of high-entropy stabilized solid-solution in multi-component alloys," Mater. Chem. Phys. **132**(2–3), 233–238 (2012).

[46] A. Takeuchi, and A. Inoue, "Classification of Bulk Metallic Glasses by Atomic Size Difference, Heat of Mixing and Period of Constituent Elements and Its Application to Characterization of the Main Alloying Element," Mater. Trans. **46**(12), 2817–2829 (2005).

[47] A. Khalajhedayati, Z. Pan, and T.J. Rupert, "Manipulating the interfacial structure of nanomaterials to achieve a unique combination of strength and ductility," Nat. Commun. **7**, (2016).

[48] J.D. Schuler, C.M. Barr, N.M. Heckman, G. Copeland, B.L. Boyce, K. Hattar, and T.J. Rupert, "In Situ High-Cycle Fatigue Reveals Importance of Grain Boundary Structure in Nanocrystalline Cu-Zr," Jom **71**(4), 1221–1232 (2019).

[49] K. Oh-Ishi, K. Edalati, H.S. Kim, K. Hono, and Z. Horita, "High-pressure torsion for enhanced atomic diffusion and promoting solid-state reactions in the aluminum-copper system," Acta Mater. **61**(9), 3482–3489 (2013).

[50] Y. Jiang, Y. Liu, H. Zhou, S. Taheriniya, B. Bian, L. Rogal, J.T. Wang, S. Divinski, and G. Wilde, "Diffusion in ultra-fine-grained CoCrFeNiMn high entropy alloy processed by equal-channel angular pressing," J. Mater. Sci. **59**(14), 5805–5817 (2024).

2020

# Hydrogen-Mediated Control of Phase Formation and Microstructure Evolution - Supplementary Information

# Figures

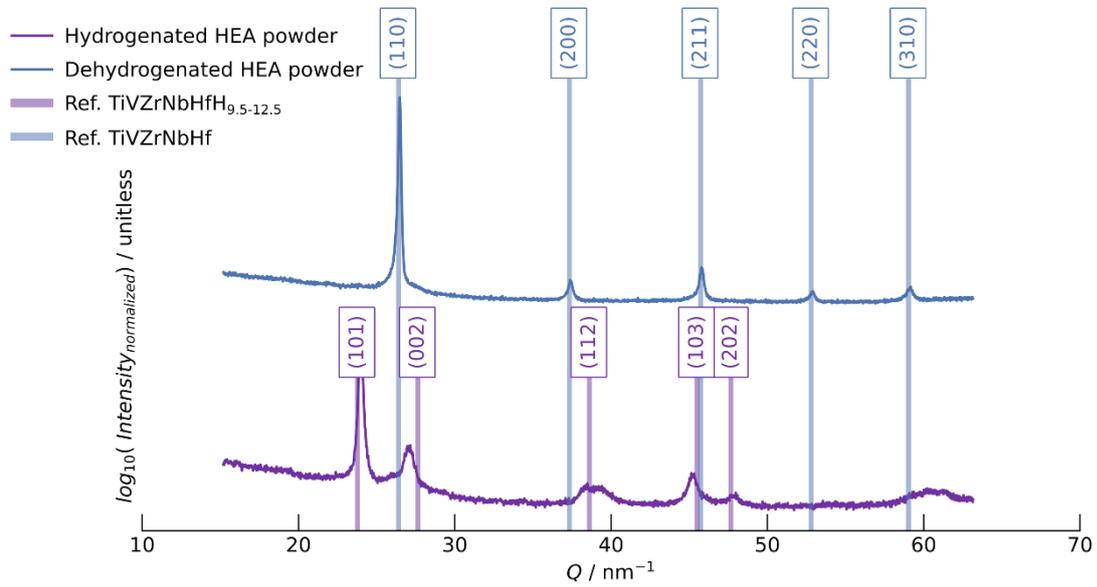

**Figure S1:** XRD patterns of the HEA hydride obtained by self-pulverization and the HEA powders obtained by subsequent annealing.

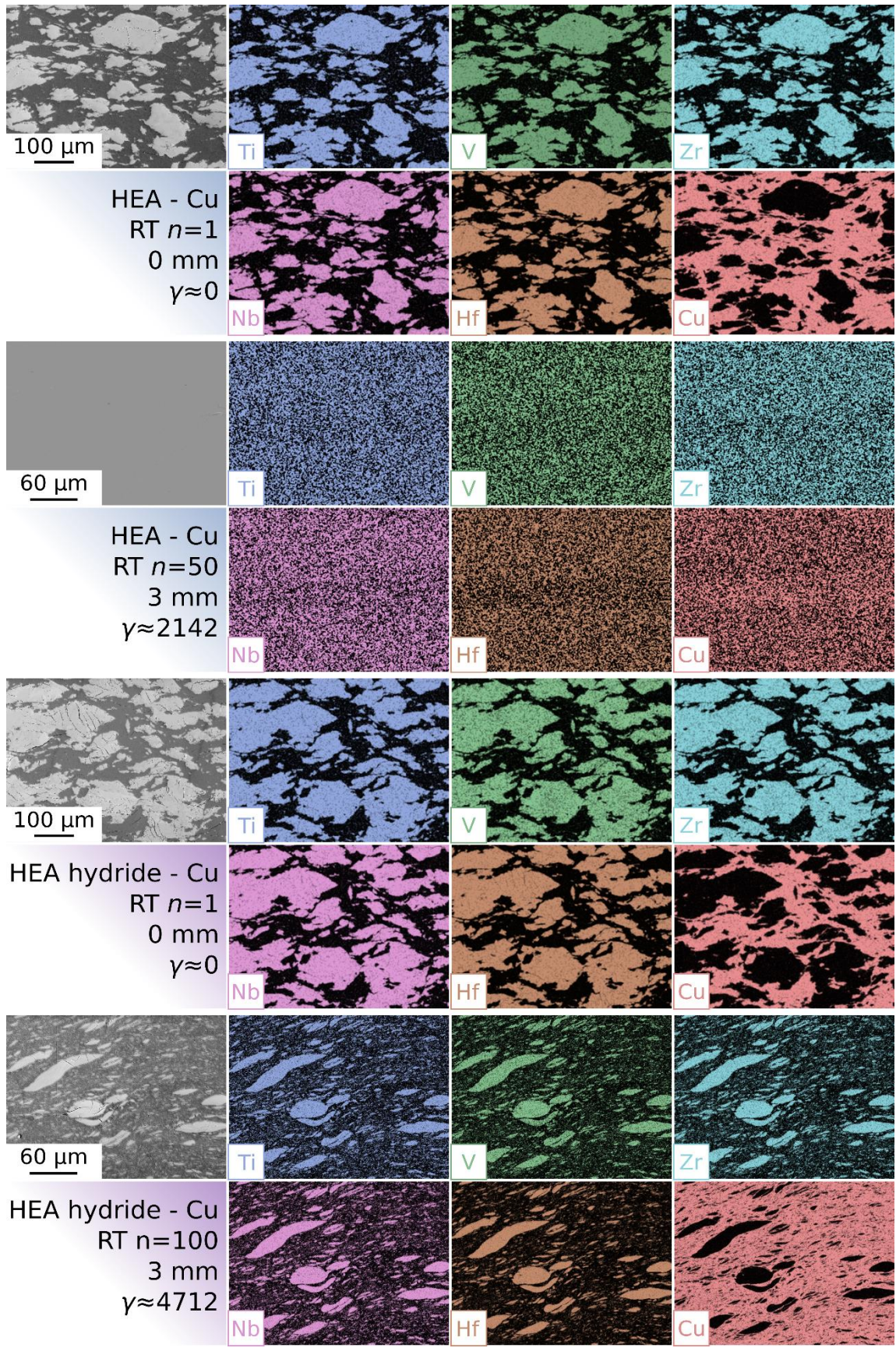

**Figure S2:** Secondary electron SEM micrographs and EDX maps of the HEA-Cu and HEA hydride-Cu composites for varying amounts of HPT deformation.

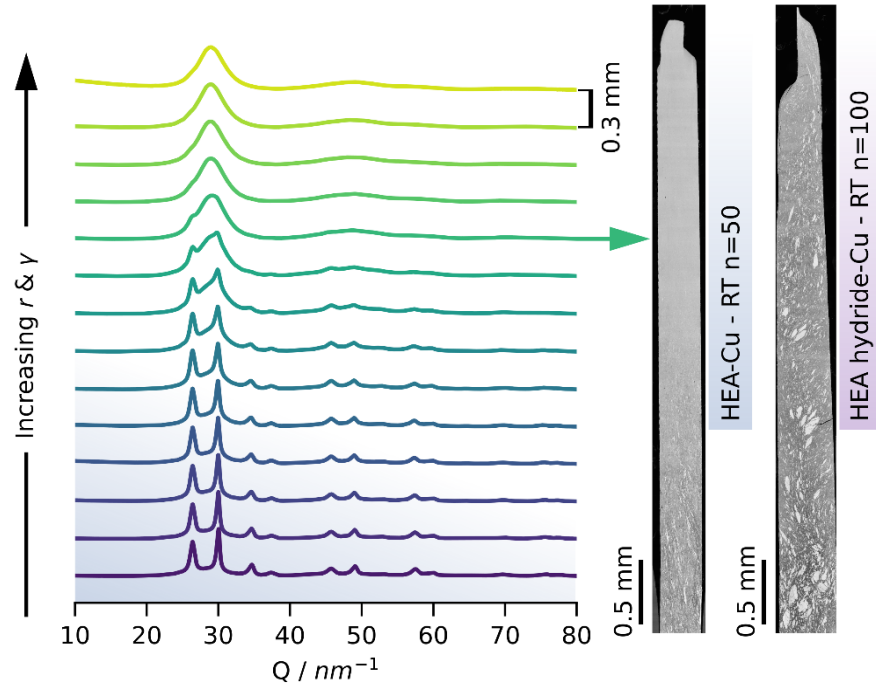

**Figure S3:** Cross-sections of the HEA-Cu and the HEA hydride-Cu disks after HPT deformation at RT and *n*=50 and 100. Radius-specific Synchrotron XRD patterns are plotted for the HEA-Cu composite at the respective positions.

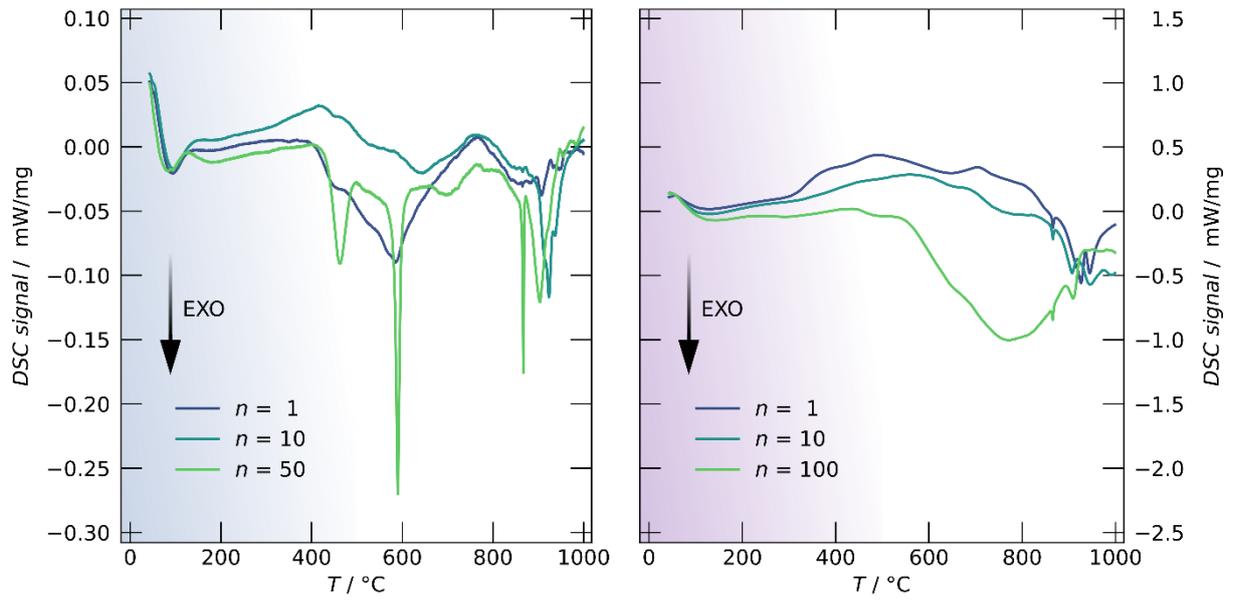

**Figure S4:** DSC measurements of the (a) HEA-Cu composite after $n$=1, 10, and 50 (10 K min$^{-1}$) and (b) the HEA hydride-Cu composite with $n$=1, 10, and 100 (20 K min$^{-1}$).

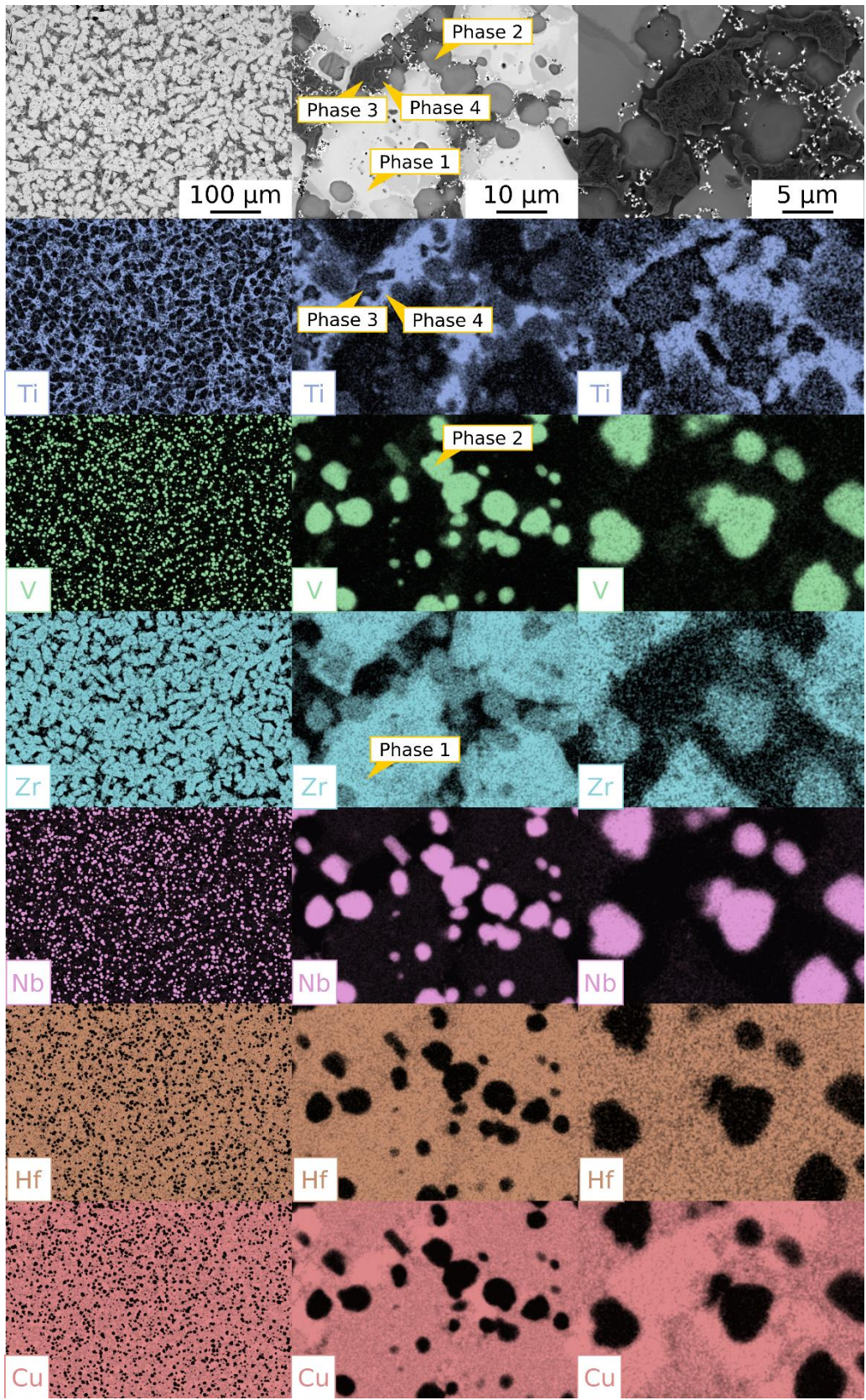

**Figure S5:** BSE SEM micrographs and associated EDX maps of the HEA-Cu composite with *n*=50 after DSC characterization (heated to 1000 °C).

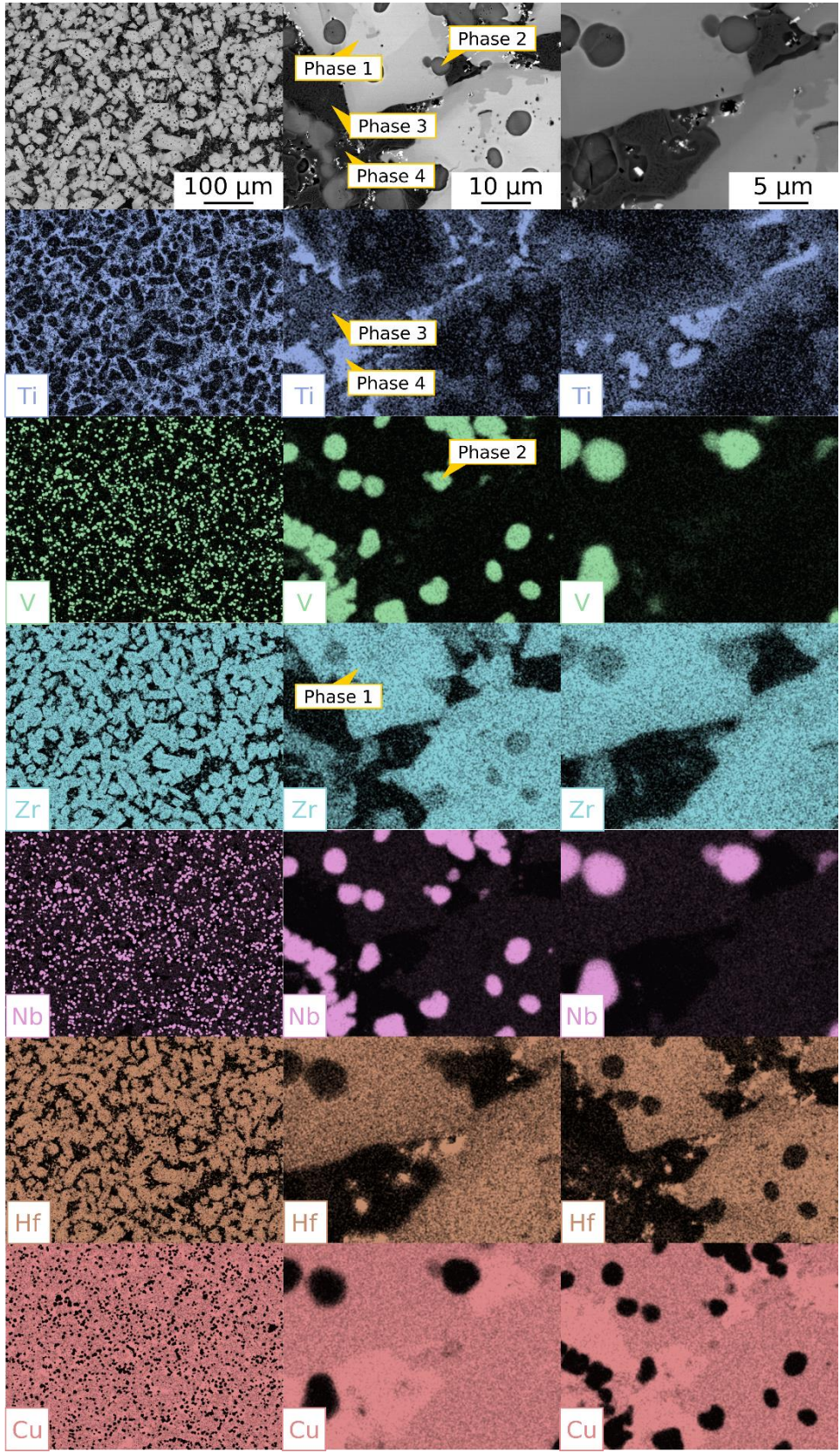

**Figure S6:** BSE SEM micrographs and associated EDX maps of the HEA hydride-Cu composite with *n*=100 after DSC characterization (heated to 1000 °C).

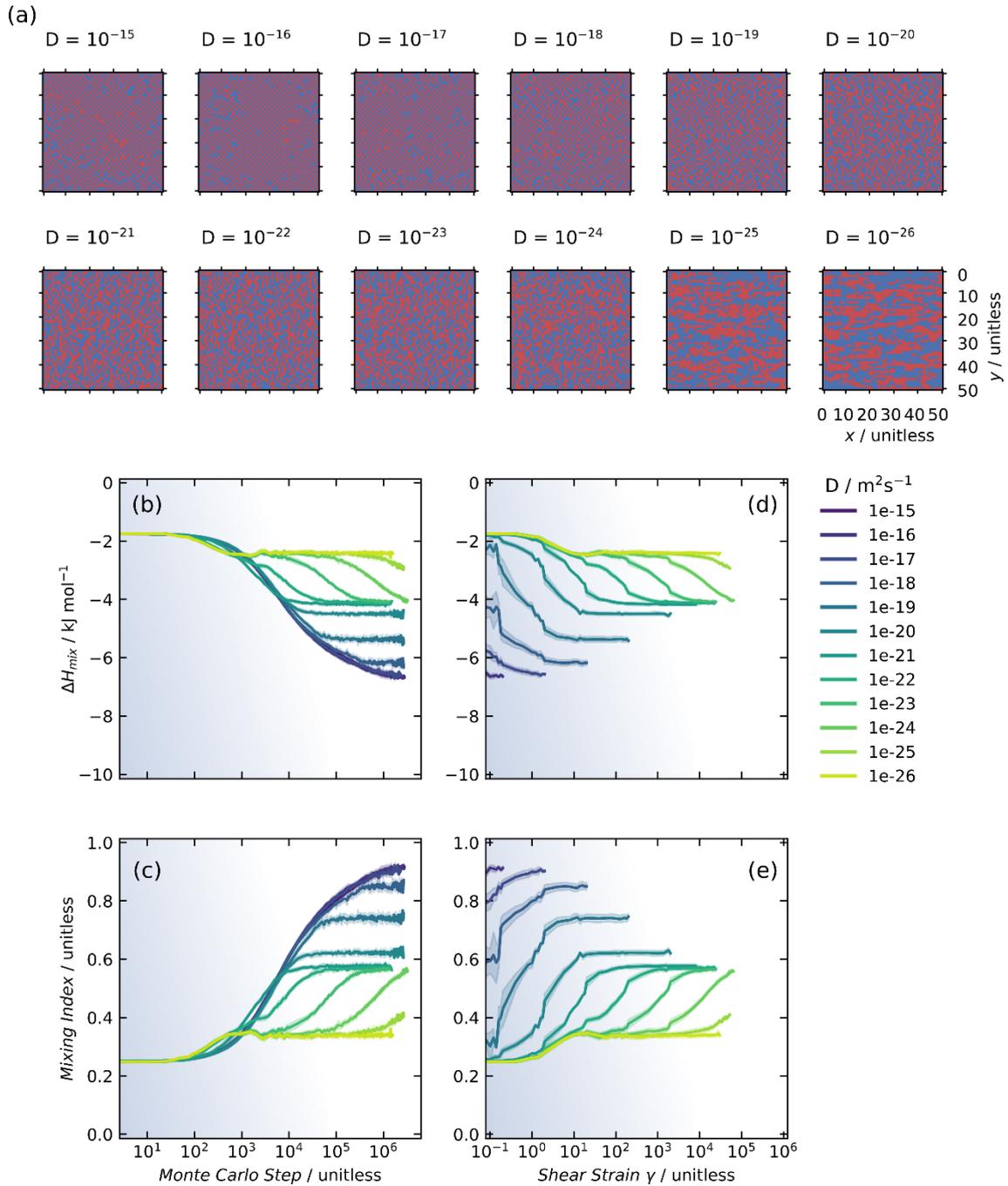

**Figure S7:** Results of the Monte Carlo simulations of the HEA-Cu composite without any hydrogen and assuming different diffusion coefficients (in m$^2$ s$^{-1}$). (a) Final grids and calculated energies and mixing indices as functions of (b,c) MC steps and (d,e) shear strain. Only horizontal shear events were allowed.

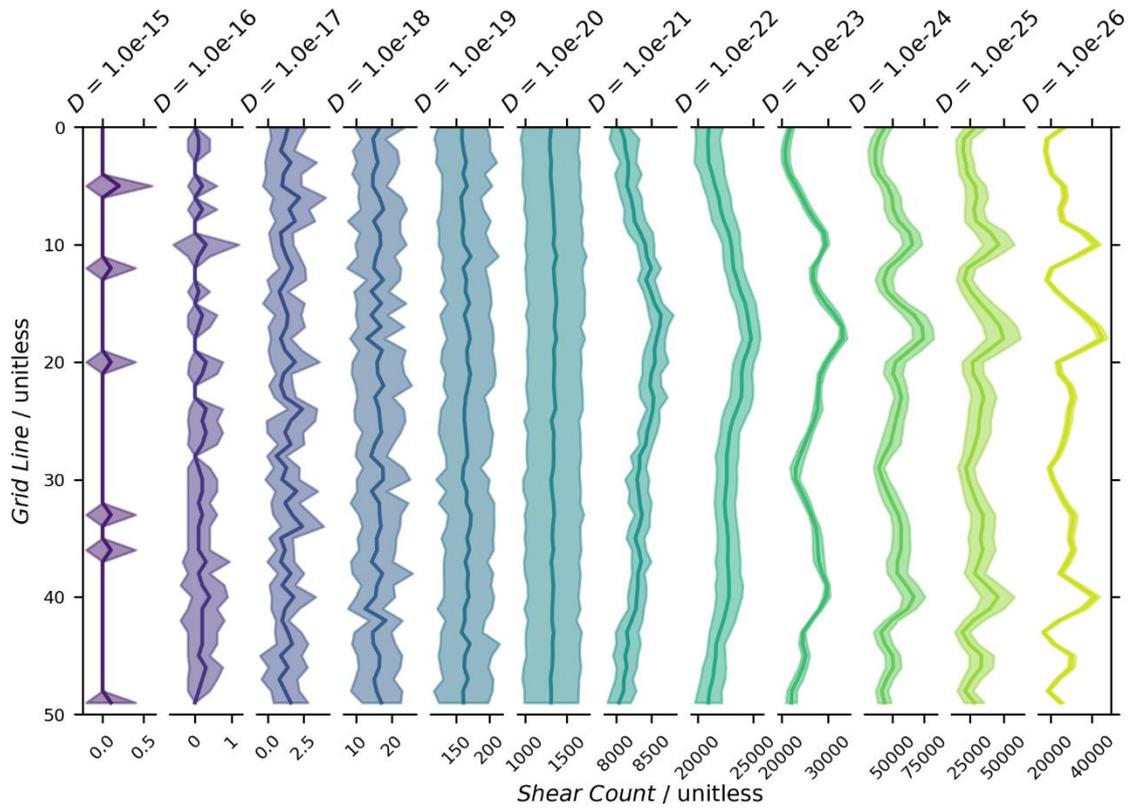

**Figure S8:** Distribution of the shear steps (horizontal only) in HEA-Cu composites derived from the Monte Carlo simulations assuming different diffusion coefficients (in m$^2$ s$^{-1}$). Only horizontal shear events were allowed.

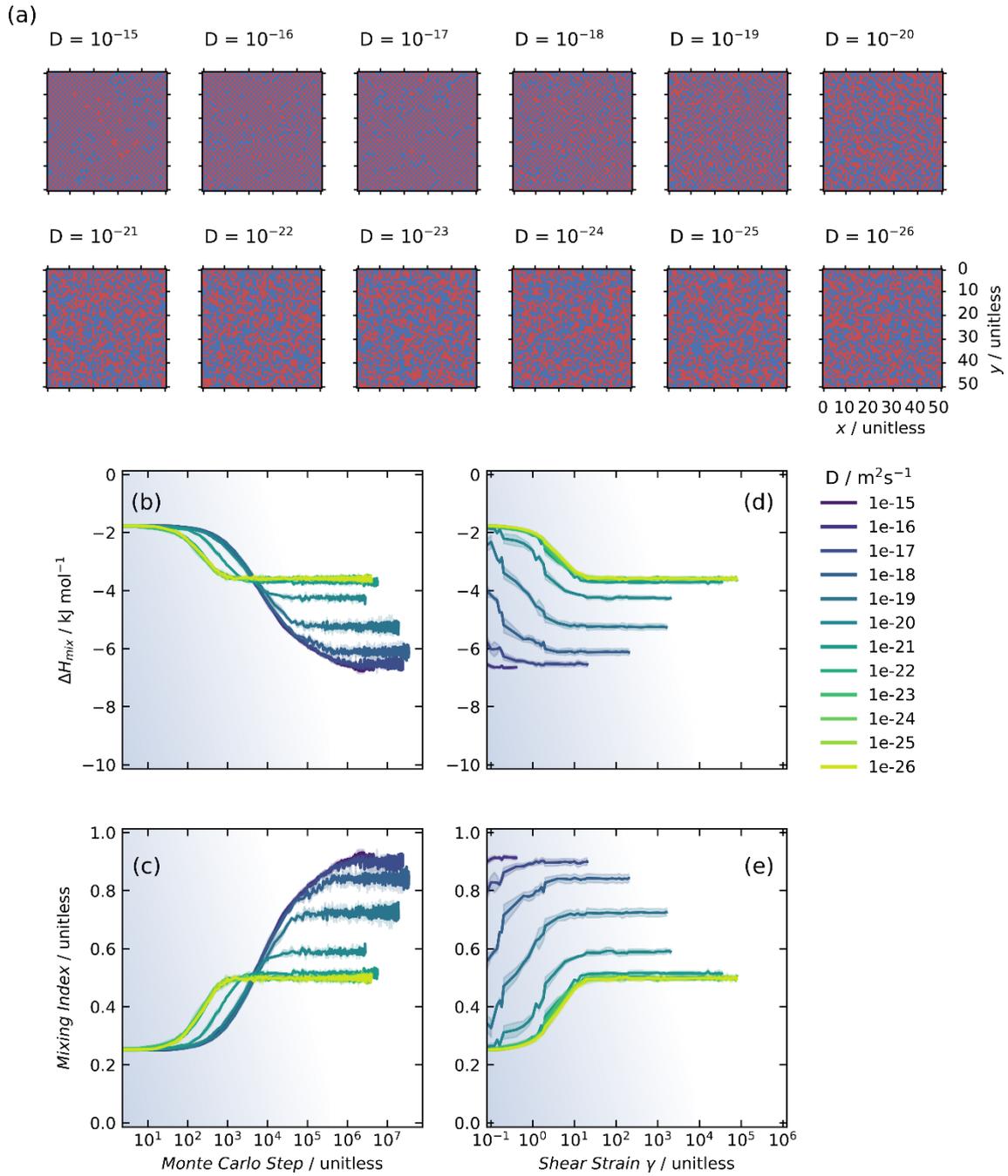

**Figure S9:** Results of the Monte Carlo simulations of the HEA-Cu composite without any hydrogen and assuming different diffusion coefficients (in m$^2$ s$^{-1}$). (a) Final grids and calculated energies and mixing indices as functions of (b,c) MC steps and (d,e) shear strain. In addition to the horizontal shear events, there was a 20 % chance for a vertical shear event.

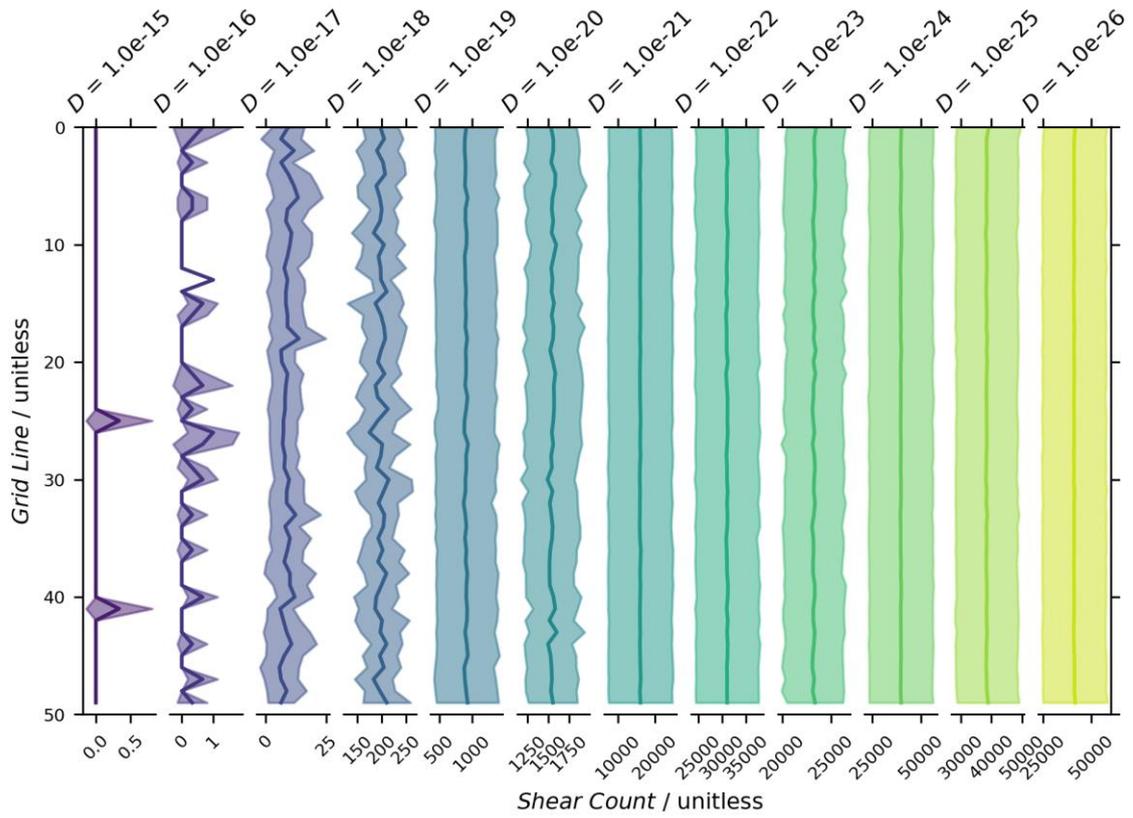

**Figure S10:** Distribution of the shear steps (horizontal only) in HEA-Cu composites derived from the Monte Carlo simulations assuming different diffusion coefficients (in m$^2$ s$^{-1}$). In addition to the horizontal shear events, there was a 20 % chance for a vertical shear event.

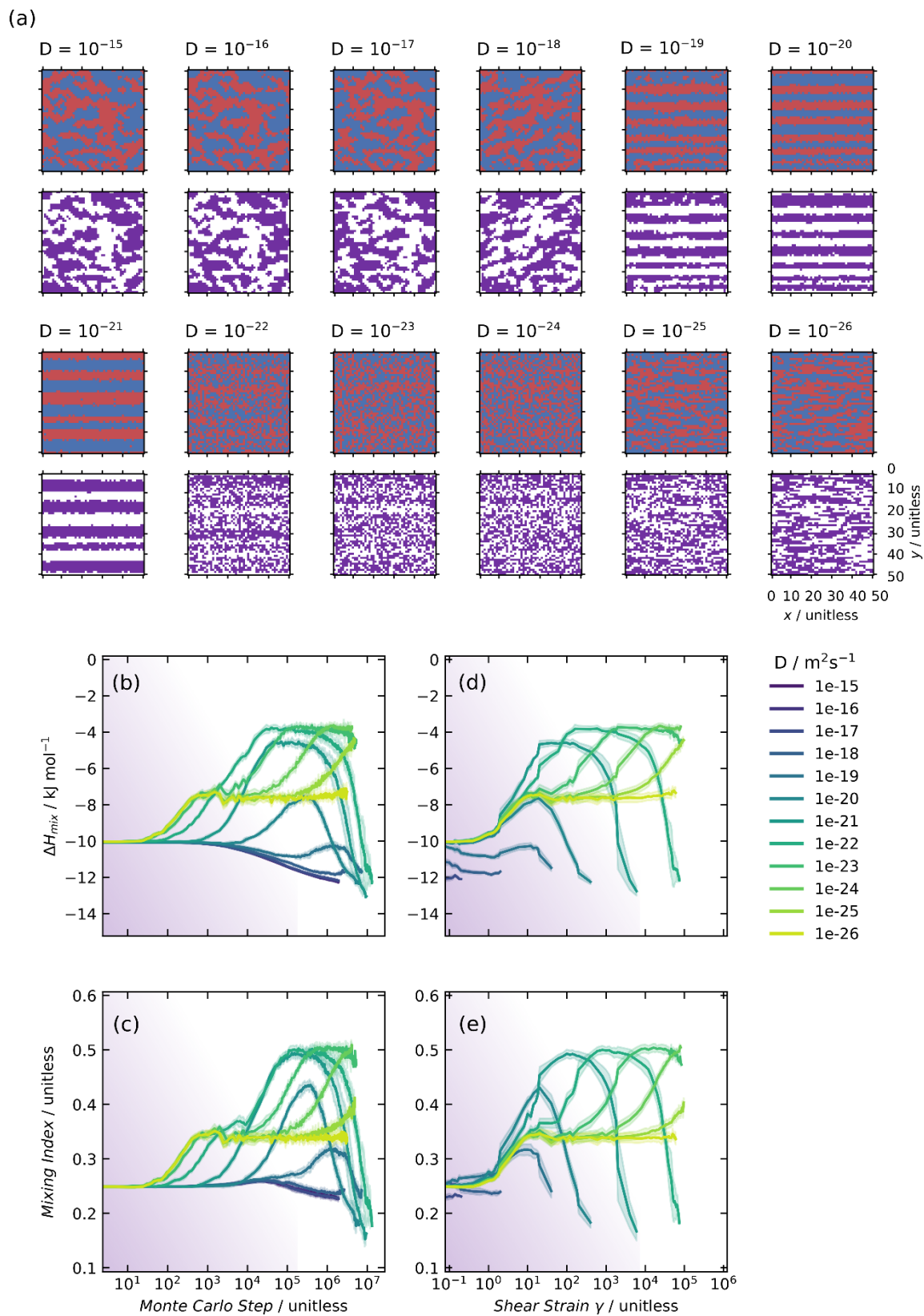

**Figure S11:** Results of the Monte Carlo simulations of the HEA hydride-Cu composite, i.e., with hydrogen, and assuming different diffusion coefficients (in m$^2$ s$^{-1}$). (a) Final grids and calculated energies and mixing indices as functions of (b,c) MC steps and (d,e) shear strain. Only horizontal shear events were allowed.

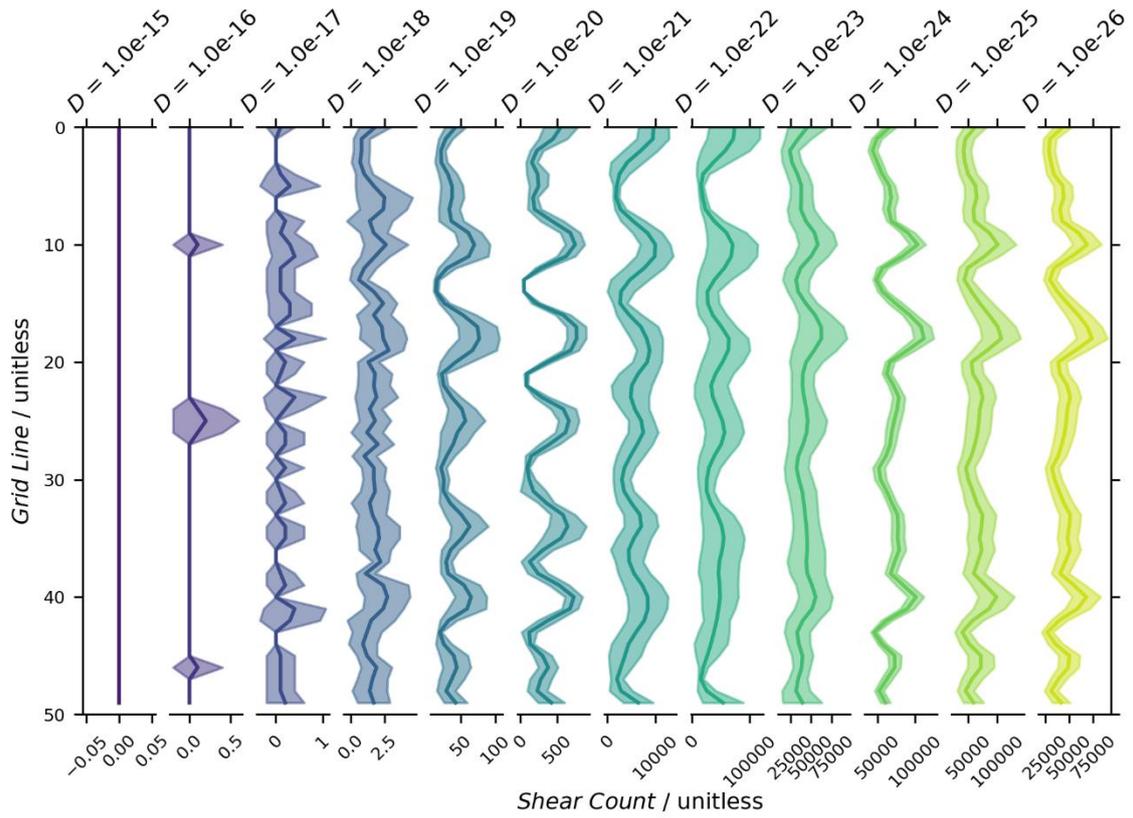

**Figure S12:** Distribution of the shear steps (horizontal only) in HEA hydride-Cu composites derived from the Monte Carlo simulations assuming different diffusion coefficients (in $m^2\ s^{-1}$). Only horizontal shear events were allowed.

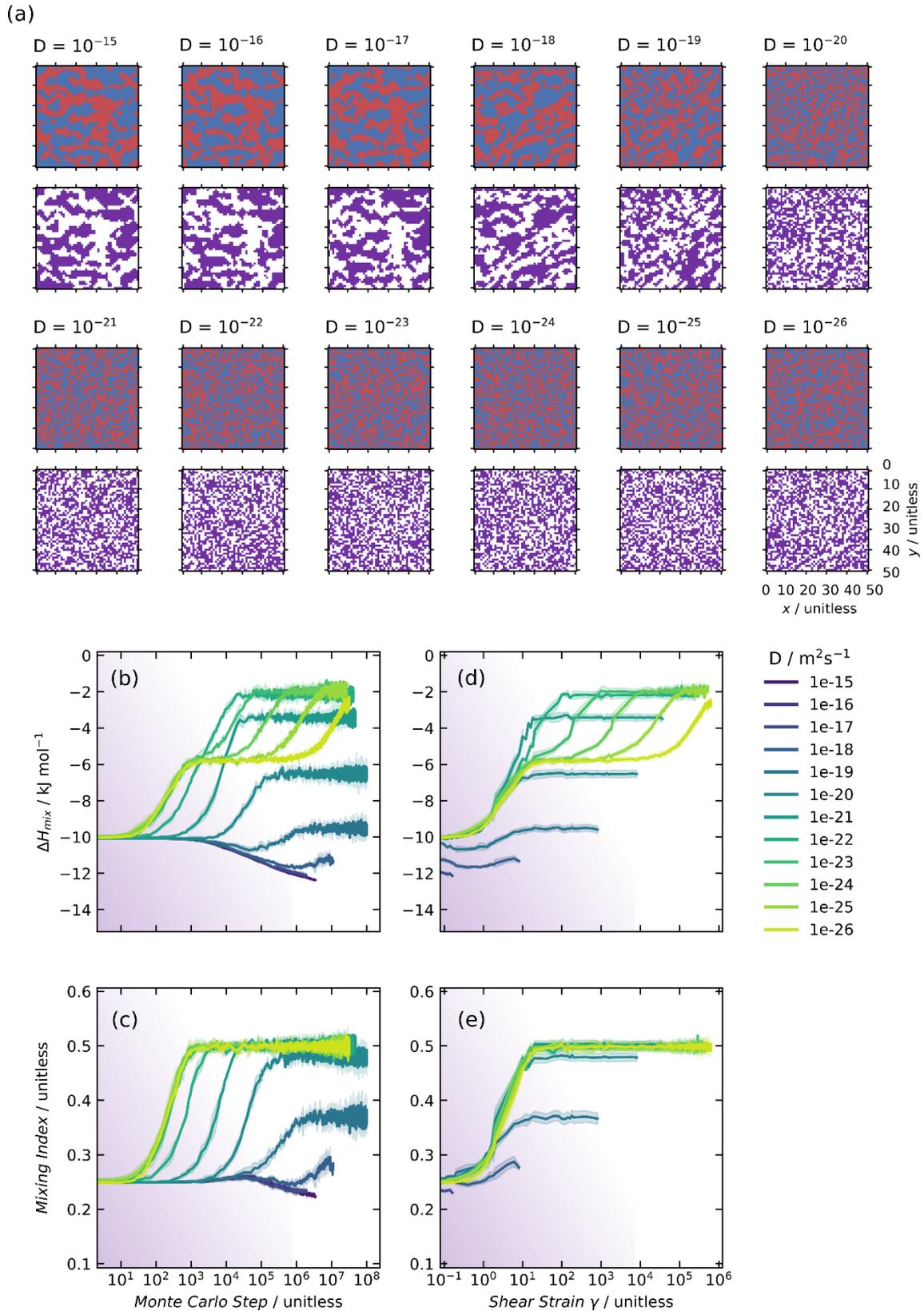

**Figure S13:** Results of the Monte Carlo simulations of the HEA hydride-Cu composite, i.e., with hydrogen, and assuming different diffusion coefficients (in m$^2$ s$^{-1}$). (a) Final grids and calculated energies and mixing indices as functions of (b,c) MC steps and (d,e) shear strain. In addition to the horizontal shear events, there was a 20 % chance for a vertical shear event.

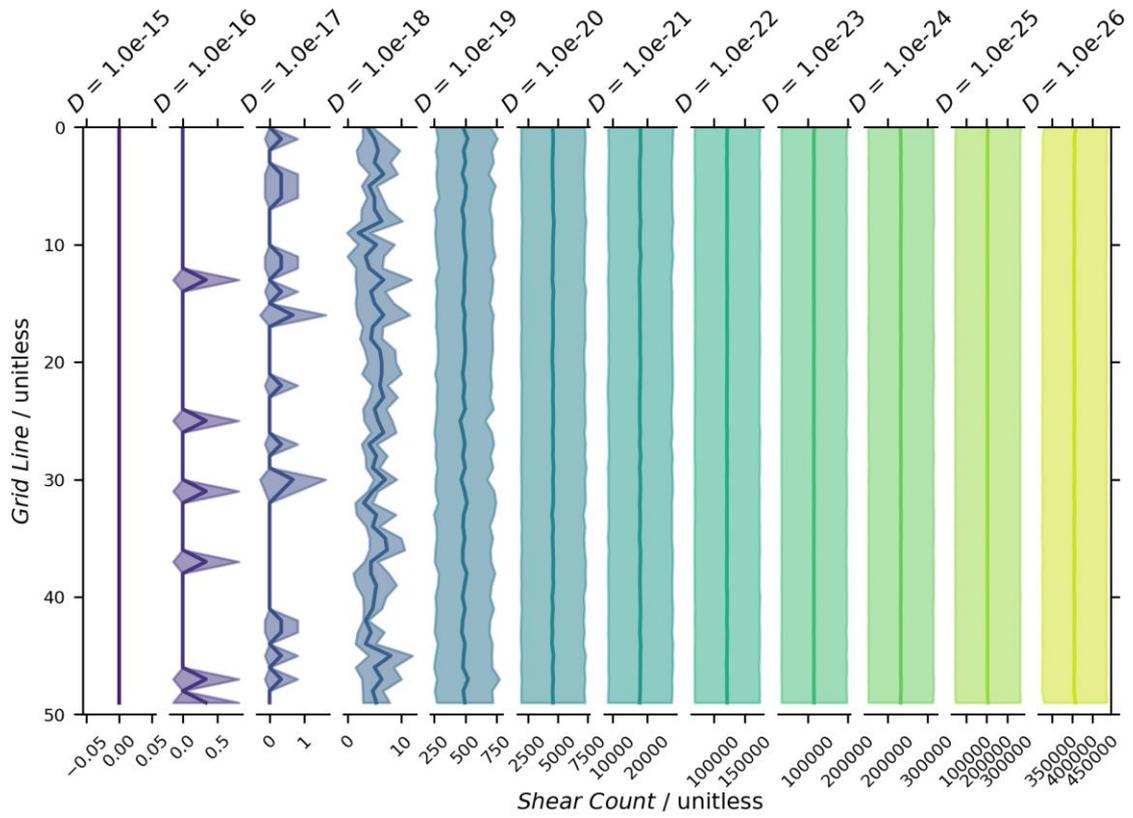

**Figure S14:** Distribution of the shear steps (horizontal only) in HEA hydride-Cu composites derived from the Monte Carlo simulations assuming different diffusion coefficients (in m$^2$ s$^{-1}$). In addition to the horizontal shear events, there was a 20 % chance for a vertical shear event.

# Tables

**Table S1:** Chemical compositions of the HEA-Cu composite obtained from EDX measurements of the initial (compacted) microstructure, as-HPTdeformed (RT, n=50, see **Figure S2**), and following DSC analysis at temperatures up to 1000 °C. The phases after DSC are indicated in **Figure S5**.

|  | Ti / at. % | V / at. % | Cu / at. % | Zr / at. % | Nb / at. % | Hf / at. % |
|---|---|---|---|---|---|---|
| **HEA - Cu - RT - n=1 - 0 mm ($\gamma \approx 0$, initial microstructure)** | | | | | | |
| HEA | 22.2 ± 0.5 | 19.6 ± 0.5 | 1.1 ± 0.6 | 19.7 ± 0.5 | 19.7 ± 0.7 | 17.7 ± 0.5 |
| Cu | 0.0 ± 0.0 | 0.0 ± 0.0 | 99.4 ± 0.2 | 0.0 ± 0.0 | 0.0 ± 0.0 | 0.4 ± 0.1 |
| **HEA - Cu - RT - n=50 - 3 mm ($\gamma \approx 2142$, as-HPT)** | | | | | | |
| HEA-Cu glass | 8.9 ± 0.5 | 8.0 ± 0.5 | 60.0 ± 2.5 | 8.0 ± 0.6 | 8.1 ± 0.6 | 7.0 ± 0.5 |
| **HEA hydride - Cu composite - RT- n=50 - Post-DSC (to 1000 °C, 10 K s$^{-1}$)** | | | | | | |
| Phase 1 (Large grains) | 4.7 ± 1.2 | 0.2 ± 0.1 | 76.0 ± 0.4 | 10.9 ± 0.8 | 0.3 ± 0.1 | 7.9 ± 0.5 |
| Phase 2 (Globular grains) | 8.0 ± 0.9 | 40.1 ± 1.9 | 6.5 ± 4.0 | 0.1 ± 0.2 | 44.9 ± 2.9 | 0.3 ± 0.3 |
| Phase 3 (Intergranular) | 7.4 ± 3.2 | 0.3 ± 0.1 | 91.4 ± 3.4 | 0.1 ± 0.2 | 0.0 ± 0.0 | 0.9 ± 0.4 |
| Phase 4 (Interphase) | 19.1 ± 3.5 | 0.7 ± 0.7 | 76.2 ± 4.7 | 1.2 ± 1.2 | 0.5 ± 0.6 | 2.3 ± 2.4 |

**Table S2:** Chemical compositions of the HEA hydride-Cu composite obtained from EDX measurements of the initial (compacted) microstructure, as-HPTdeformed (RT, n=100, see **Figure S2**), and following DSC analysis at temperatures up to 1000 °C. The phases after DSC are indicated in **Figure S6**.

|  | Ti / at. % | V / at. % | Cu / at. % | Zr / at. % | Nb / at. % | Hf / at. % |
|---|---|---|---|---|---|---|
| **HEA hydride - Cu composite - RT- n=1 – 0 mm ($\gamma \approx 0$, initial microstructure)** | | | | | | |
| HEA hydride | 22.2±0.5 | 20.2±1.7 | 1.1±0.5 | 19.5±0.8 | 19.7±1.7 | 17.3±0.7 |
| Cu | 0.0±0.0 | 0.1±0.0 | 99.4±0.2 | 0.0±0.0 | 0.0±0.0 | 0.5±0.2 |
| **HEA hydride - Cu composite - RT- n=100 – 3 mm ($\gamma \approx 4712$, as-HPT)** | | | | | | |
| HEA hydride | 21.9±0.4 | 19.4±0.8 | 2.5±0.7 | 19.6±0.5 | 19.7±0.6 | 16.9±0.4 |
| (Nano)composite | 6.4±1.6 | 5.7±1.4 | 71.5±6.7 | 5.5±1.3 | 5.7±1.4 | 5.2±1.1 |
| **HEA hydride - Cu composite - RT- n=100 - Post-DSC (to 1000 °C, 20 K s$^{-1}$)** | | | | | | |
| Phase 1 (Large grains) | 4.9±1.9 | 0.1±0.1 | 76.1±0.6 | 11.8±0.8 | 0.3±0.1 | 6.7±0.9 |
| Phase 2 (Globular grains) | 7.4±0.9 | 42.4±0.8 | 3.8±0.3 | 0.0±0.1 | 46.3±0.8 | 0.1±0.1 |
| Phase 3 (Intergranular) | 8.0±1.0 | 0.3±0.1 | 91.1±1.1 | 0.0±0.0 | 0.0±0.0 | 0.5±0.1 |
| Phase 4 (Interphase) | 18.9±4.7 | 1.1±1.0 | 75.8±5.5 | 1.3±0.8 | 1.2±1.3 | 1.7±0.8 |

**Table S3:** $\Delta H_{AB}^{mix}$ of all elemental pairs in kJ mol$^{-1}$ taken from Takeuchi and Inoue.[2]

|  | Ti | V | Zr | Nb | Hf | Cu |
|---|---|---|---|---|---|---|
| Ti | - | -2 | 0 | 2 | 0 | -9 |
| V | - | - | -4 | -1 | -2 | 5 |
| Zr | - | - | - | 4 | 0 | -23 |
| Nb | - | - | - | - | 4 | 3 |
| Hf | - | - | - | - | - | -17 |
| Cu | - | - | - | - | - | - |

**Table S4:** $\Delta H_{mix}$, $\Delta S_{mix}$, and $\Delta G_{mix}$ of the various metal-metal and metal-hydrogen interactions in the TiVZrNbHf-Cu-H systems computed using the enthalpy values in **Table S3** and **Equations S1-S4**.[2–6] The $\Delta H_{mix}$ values per atom were used for the Monte Carlo simulations.

| Type of interaction | $\Delta H_{mix}$ / kJ mol$^{-1}$(H) | $\Delta S_{mix}$ / J K$^{-1}$ mol$^{-1}$ (H) | $\Delta G_{mix}$ (300 K) / kJ mol$^{-1}$ | $\Delta H_{mix}$ / kJ atom$^{-1}$ |
|---|---|---|---|---|
| Hea-Cu (A-B) | −7.26 | 9.75 | −10.19 | −1.21·10$^{−23}$ |
| Hea (A-A) | 0.16 | 13.38 | −3.85 | 2.66·10$^{−25}$ |
| Cu (B-B) | 0.00 | 0.00 | 0.00 | 0.00 |
| Hea-H (A-H) | −29.50 | −41.00 | −17.20 | −4.90·10$^{−23}$ |
| Cu-H (B-H) | 42.45 | −50.48 | 57.60 | 7.05·10$^{−23}$ |

# Text

**Calorimetric characterization**

DSC measurements (heating rate of 10 and 20 K min$^{-1}$) were conducted to understand better the metastable character of the materials prepared in this study. **Figure S4** depicts the DSC curves of the HEA-Cu and HEA hydride-Cu composites.

As seen in **Figure S4 (a)** for the HEA-Cu composites, after *n*=1 and n=10, two relatively broad peaks are visible at 500 and 600 °C, while at *n*=50, these become two pronounced and well-defined peaks. These are associated with the decomposition of the solid-solution/amorphous phase. SEM and EDX investigations of the DSC samples, given in **Figure S5,** show decomposition into a complex multiphase material containing a Cu-Zr-rich phase (Phase 1, large grains in **Table S1**), a Nb-V-rich phase (Phase 2, globular grains in **Table S1**) and a Cu-rich intergranular phase (Phase 3, Intergranular in **Table S1**) surrounded by a Ti-enriched interphase (Phase 4, Interphase in **Table S1**). The exact chemical compositions are given in **Table S1.**

The DSC curves of the less deformed hydride composites with *n*=1 and 10 given in **Figure S4 (b),** exhibit three convoluted endothermic peaks related to dehydrogenation.[1] The peaks become less pronounced and shift to higher temperatures with increasing strain. This suggests that embedding the HEA hydride in a rigid Cu matrix stabilizes the hydride, which then desorbs hydrogen only at higher temperatures. No distinct endothermic signal was obtained for the most severely deformed sample at *n*=100. On the contrary, broad exothermic peaks were observed above temperatures of about 500 °C. Although SPD could hinder the hydride decomposition, it is unlikely that it is stable up to 1000 °C. The HPT-deformed nanocrystalline material is in a highly metastable state. Upon heating to high temperatures, interdiffusion and associated phase transformations will occur as soon as the stabilizing effect of the hydrogen/hydride is removed. Such exothermic processes could surpass and mask the endothermic hydrogen desorption. This highlights the stabilizing character of hydrogen for this particular nanostructure. The microstructure after DSC measurements is given in **Figure S6,** showing a similar microstructure as the HEA-Cu composite without hydrogen.

**Mechanical characterization**

Hardness measurements of the HEA and Cu reveal distinct strength differences. The HEA registers 6.1 ± 0.8 GPa, while the HEA hydride is slightly stronger at 6.8 ± 0.5 GPa. In contrast, Cu is significantly softer at 1.4 ± 0.1 GPa. These results indicate that the hardness differences between the HEA / HEA hydride and Cu are consistent in the hydrogenated and dehydrogenated states. It should be noted that cracking at the indents was observed, especially for the hydrides, so the reported hardness values are semi-quantitative estimations. Therefore, a strength difference of 5 was assumed for both the HEA and HEA hydride systems in the Monte Carlo simulation; see below for details.

**Thermodynamic Considerations & Monte Carlo Simulations**

The thermodynamic considerations are based on the work of Yang and Zhang on predicting the stability of HEAs.[3] Using the enthalpies of mixing of the respective binary liquid alloys, $\Delta H_{AB}^{mix}$, given in **Table S1**, the regular solution interaction parameter $\Omega_{ij}$ was calculated using

$$\Omega_{ij} = 4\,\Delta H_{AB}^{mix}. \tag{S1}$$

Based on this, the enthalpy of mixing of a multi-component alloy can be estimated from

$$\Delta H_{mix} = \sum_{i=1, i \neq j}^{n} \Omega_{ij}\, c_i\, c_j, \tag{S2}$$

with $c_i$ and $c_j$ being the atomic percent of the alloy system. The entropy of mixing $\Delta S_{mix}$ was estimated based on

$$\Delta S_{mix} = -R \sum_{i=1}^{n} (c_i \ln c_i), \tag{S3}$$

with $R$ being the ideal gas constant, i.e., 8.314 J K$^{-1}$ mol$^{-1}$. From the enthalpies and entropies, the Gibb's free energy associated with alloy formation and, as an extension, mechanical alloying can be evaluated using

$$\Delta G_{mix} = \Delta H_{mix} - T\,\Delta S_{mix} \tag{S4}$$

with $T$ being the deformation temperature during HPT, i.e., 300 K. The values for $\Delta H_{mix}$, $\Delta S_{mix}$, and $\Delta G_{mix}$ calculated using this methodology are provided in **Table S4** for TiVZrNbHf and TiVZrNbHf-Cu,[2,3] together with the respective hydrogen interactions, i.e., Cu-H and TiVZrNbHf-H.[4–6]

A model based on Monte Carlo simulations was developed to validate the impact of hydrogen and varying hydrogen affinities on the microstructural evolution during HPT. Simulations were conducted under conditions with and without hydrogen to compare the respective effects.

A (50x50) lattice representing the atom positions and occupied with A atoms (HEA - high hydrogen affinity) and B atoms (Cu - low hydrogen affinity) was defined in an initial step. Interactions were limited to nearest neighbors, i.e., each atom interacted with the four surrounding lattice sites. Based on the mixing enthalpies $\Delta H_{mix}$, calculated using **Equation S2** and given in **Table S4**, the A-A and A-B interaction energies were estimated with $2.66 \cdot 10^{-25}$ kJ per atom and $-1.21 \cdot 10^{-23}$ kJ per atom, respectively. B-B interactions (Cu-Cu) were set to 0.

Each Monte Carlo (MC) step included a defined probability for diffusion and shear, respectively. The diffusion step swaps two randomly chosen neighboring atoms. Shear results in the movement by one lattice position along a randomly selected horizontal or vertical line in the grid to mimic the HPT deformation. Periodic boundary conditions were imposed on both diffusion and shear steps.

The probability of shear or diffusion attempts was calculated based on the strain rate and diffusion coefficient during HPT. Notably, the diffusion coefficient can change significantly in SPD-deformed materials compared to the bulk coarse-grained material, and could therefore only be estimated in this study. Therefore, it was varied in a wide range from $10^{-26}$ to $10^{-15}$ m$^2$ s$^{-1}$ to get a complete picture, with realistic values most likely in the range of $10^{-18}$-$10^{-20}$ m$^2$ s$^{-1}$.[7–11] Based on the diffusion coefficient, the jump frequency $\Gamma$ can be calculated using

$$\Gamma = \frac{4 D}{\lambda^2} \quad (S5)$$

With $\lambda$ being the jump distance (estimated as 2.9 Å), $D$ the diffusion coefficient, and the factor of 4 accounting for the 2D grid geometry.[12] This frequency should not be confused with the attempt frequency $\nu$ ($\approx 10^{13}$) often referred to in diffusion, as a MC attempt already represents a successful jump surmounting the initial activation barrier.

At 1.27 rpm, a radius of 4 mm, and a disk thickness of 0.5 mm, a shear strain $\gamma$ of 1.06 is achieved within 1 s. In the MC simulations with a grid of 50x50, a shear strain of 1 is equivalent to 50 shear steps. Consequently, 53.2 shear steps per second result in this shear strain rate. The resulting jump frequencies $\Gamma$ and shear rates yield the respective (relative) probabilities of diffusion and shear attempts in the system. Two types of simulations were performed, allowing only horizontal shear or giving a 20 % chance of a vertical shear event, respectively. The latter mimics less ideal shear conditions or turbulent plastic flow during HPT.[13]

Additionally, each diffusion attempt was accepted or declined based on the Metropolis criterion,[14] with the temperature set to 300 K and respective energy changes calculated using the $\Delta H_{mix}$ in **Table S4**.

Each shear attempt was accepted with a probability linearly scaled by the types of atoms present along the shear line. This simulated the flow strength differences between Cu and the HEA and allowed us to model deformation localization and its interaction with mechanical alloying/diffusion. However, unlike diffusion, each shear attempt is repeated until a shear event occurs, i.e., a particular shear strain rate is enforced. Both HEA and HEA hydride were assumed to have five times the flow strength of the Cu phase, which is in line with experimental observations based on microhardness measurements mentioned above.

The total energy was calculated by summing up all nearest-neighbor interactions. The ordering/mixing was assessed by examining the types of neighbors for each atom. A *mixing index* $i_{mix}$ was defined as

$$i_{mix} = \frac{1}{4M^2}\sum_{i=1}^{M}\sum_{j=1}^{M}\sum_{k \in N(i,j)}(1 - \delta(G(i,j), G(k))) \qquad (S6)$$

With $M$ being the grid size along the two dimensions, $G(i,j)$ the atom at position $(i,j)$, $N(i,j)$ represents the four neighbors of every site, accounting for periodic boundary conditions. A value of 0 indicates no A-B pairs, 0.5 is an even distribution with 50% A-B and 50% A-A/B-B pairs and 1 exclusively A-B pairs. Values 0 and 1 correspond to highly ordered states, though distinctly different, while 0.5 signified a random atom distribution.

Hydrogen was included by defining a second, superimposed sub-lattice. Hydrogen was positioned at the position of the A atoms to reflect the hydride. The A-H and B-H interactions were estimated based on the enthalpy of the hydride formation of TiVZrNbHf and the enthalpy of hydrogen dissolution in Cu. Using the values in **Table S4**,[4–6] the former amounts to $-4.90 \cdot 10^{-23}$ kJ atom$^{-1}$, and the latter to $7.05 \cdot 10^{-23}$ kJ atom$^{-1}$.

The hydrogen diffusivity was estimated to be significantly faster than the metal diffusion[15] and was set to 10 times the latter's value. A even higher ratio was avoided to reduce the computational demand of the MC simulation.

The in-depth results of the MC simulations for the HEA-Cu and HEA hydride-Cu systems are presented in **Figures S7** to **S14**.

**Figures 7-8** show the HEA-Cu results for horizontal shear only, while **Figures 9-10** show the results for horizontal and vertical shear. **Figure S7 (a)** shows representative grids for different diffusion coefficients $D$, while the evolution of calculated energies and mixing indices is illustrated as functions of **(b,c)** MC steps and **(d,e)** shear strain. The simulations demonstrate convergence, with consistent results across multiple MC runs. Large diffusion coefficients result in significant mixing and an ordered structure, reflected by a mixing index near 1. Conversely, at lower diffusion coefficients, where shear probability increases, the mixing index approaches 0.5, indicating a random solid solution. **Figure S8** reveals a uniform distribution of shear events across the grid, showing that interdiffusion mitigates the initial tendency for shear localization and promotes a more homogeneous strain distribution. The same trends are visible in **Figures 9-10**, although these results highlight the more efficient mixing and more evenly distributed plastic deformation induced by the multi-directional shear.

**Figures 11-12** show the HEA hydride-Cu results for horizontal shear only, while **Figures 13-14** show the results for horizontal and vertical shear. **Figure S11 (a)** presents metal and hydrogen sub-grids at varying diffusivities for the HEA hydride-Cu system. At high diffusivities (low shear probability), phase separation persists with a coarse structure. As shear probability increases (lower $D$), the microstructure becomes lamellar, reflecting a strong tendency for phase separation even under high shear rates. **Figure S11 (b,c)** shows again the evolution of calculated energies and mixing indices as functions of **(b)** MC steps and **(c)** shear strain. **Figure S12** highlights pronounced strain localization in this regime. Mechanical intermixing dominates at very low diffusivities and high shear probabilities, leading to a higher mixing index. As seen in **Figure S12**, strain localization is significantly reduced in this range. Again, the same trends are visible in **Figures 13-14**, although these results highlight the more efficient mixing and more evenly distributed plastic deformation induced by the multi-directional shear.

1. Sahlberg, M., Karlsson, D., Zlotea, C. & Jansson, U. Superior hydrogen storage in high entropy alloys. *Sci. Rep.* **6**, 1–6 (2016).

2. Takeuchi, A. & Inoue, A. Classification of Bulk Metallic Glasses by Atomic Size Difference, Heat of Mixing and Period of Constituent Elements and Its Application to Characterization of the Main Alloying Element. *Mater. Trans.* **46**, 2817–2829 (2005).

3. Yang, X. & Zhang, Y. Prediction of high-entropy stabilized solid-solution in multi-component alloys. *Mater. Chem. Phys.* **132**, 233–238 (2012).

4. Griessen, R. & Riesterer, T. Heat of Formation Models. in *Hydrogen in Intermetallic Compounds I* (ed. Schlapbach, L.) 219–284 (Springer-Verlag Berlin Heidelberg, 1988).

5. Fukai, Y. *The Metal-Hydrogen System*. vol. 21 (Springer Berlin Heidelberg, 2005).

6. Karlsson, D. *et al.* Structure and Hydrogenation Properties of a HfNbTiVZr High-Entropy Alloy. *Inorg. Chem.* **57**, 2103–2110 (2018).

7. Oh-Ishi, K., Edalati, K., Kim, H. S., Hono, K. & Horita, Z. High-pressure torsion for enhanced atomic diffusion and promoting solid-state reactions in the aluminum-copper system. *Acta Mater.* **61**, 3482–3489 (2013).

8. Alhamidi, A., Edalati, K., Iwaoka, H. & Horita, Z. Effect of temperature on solid-state formation of bulk nanograined intermetallic Al3Ni during high-pressure torsion. *Philos. Mag.* **94**, 876–887 (2014).

9. Jiang, Y. *et al.* Diffusion in ultra-fine-grained CoCrFeNiMn high entropy alloy processed by equal-channel angular pressing. *J. Mater. Sci.* **59**, 5805–5817 (2024).

10. Divinski, S. V., Reglitz, G., Rösner, H., Estrin, Y. & Wilde, G. Ultra-fast diffusion channels in pure Ni severely deformed by equal-channel angular pressing. *Acta Mater.* **59**, 1974–1985 (2011).

11. Amouyal, Y., Divinski, S. V., Estrin, Y. & Rabkin, E. Short-circuit diffusion in an ultrafine-grained copper–zirconium alloy produced by equal channel angular pressing. *Acta Mater.* **55**, 5968–5979 (2007).

12. Mehrer, H. *Diffusion in Solids*. vol. 155 (Springer Berlin Heidelberg, 2007).

13. Beygelzimer, Y., Filippov, A. & Estrin, Y. 'Turbulent' shear flow of solids under high-pressure torsion. *Philos. Mag.* **103**, 1017–1028 (2023).

14. Metropolis, N., Rosenbluth, A. W., Rosenbluth, M. N., Teller, A. H. & Teller, E. Equation of State Calculations by Fast Computing Machines. *J. Chem. Phys.* **21**, 1087–1092 (1953).

15. Völkl, J. & Alefeld, G. Diffusion of hydrogen in metals. in *Hydrogen in Metals I. Topics in Applied Physics, vol 28* (eds. Alefeld, G. & Völkl, J.) 321–348 (Springer Berlin Heidelberg, 1978). doi:10.1007/3540087052_51.